\documentclass[english,12pt,aps,prd,a4paper,preprintnumbers,floatfix,nofootinbib,showpacs,superscriptaddress, notitlepage]{revtex4-1} 
\usepackage[mode=buildnew]{standalone}
\usepackage[usenames,dvipsnames]{color} 
\usepackage{graphicx}
\usepackage{bm}
\usepackage{dcolumn}
\usepackage[colorlinks=true,citecolor=darkred,urlcolor=darkred, pdfborder={0 0 0}]{hyperref}
\usepackage{setspace}
\usepackage{caption}
\usepackage{subcaption}
\usepackage{slashed}
\usepackage[normalem]{ulem}
\usepackage{float}
\usepackage[usenames,dvipsnames]{xcolor}
\usepackage{microtype}
\usepackage{lipsum}
\usepackage{amsmath}
\usepackage{amssymb}
\usepackage{mathrsfs}
\usepackage{placeins}
\usepackage{lettrine}
\input Eileen.fd

\definecolor{darkred}{rgb}{0.6,0,0}

\definecolor{linkcolor}{rgb}{0,0,0.5}
\usepackage[T1]{fontenc} 
\usepackage[compat=1.1.0]{tikz-feynhand}
\usepackage{tikz-feynman}
\tikzfeynmanset{compat=1.1.0}
\usepackage{feynmp}
\usepackage{tikzsymbols}
\usepackage{array}
\usepackage{pifont}

\definecolor{linkcolor}{rgb}{0,0,0.5}
\usepackage{orcidlink}
\usepackage{multirow}
\begin{document}
\title{\boldmath \color{BrickRed}
Flavor Imprints on Novel Low Mass Dark Matter}
\author{Ranjeet Kumar}
\email{ranjeet20@iiserb.ac.in}
\affiliation{Department of Physics, Indian Institute of Science Education and Research - Bhopal \\ Bhopal ByPass Road, Bhauri, Bhopal 462066, India}
\author{Hemant Kumar Prajapati}
\email{hemant19@iiserb.ac.in}
\affiliation{Department of Physics, Indian Institute of Science Education and Research - Bhopal \\ Bhopal ByPass Road, Bhauri, Bhopal 462066, India}
\author{Rahul Srivastava}
\email{rahul@iiserb.ac.in}
\affiliation{Department of Physics, Indian Institute of Science Education and Research - Bhopal \\ Bhopal ByPass Road, Bhauri, Bhopal 462066, India}
\author{Sushant Yadav}
\email{sushant20@iiserb.ac.in}
\affiliation{Department of Physics, Indian Institute of Science Education and Research - Bhopal \\ Bhopal ByPass Road, Bhauri, Bhopal 462066, India}
\begin{abstract}
  \vspace{1cm} 

  \noindent
We present a Majorana scotogenic-like loop framework in which neutrino mass generation and dark matter stability are intrinsically connected to the breaking of the discrete flavor symmetry $A_4$. This breaking leads to the emergence of the scoto-seesaw mechanism and a $Z_2$ symmetry. This naturally explains the solar and atmospheric mass-squared differences, $\Delta m_{\rm{sol}}^{2}$ and $\Delta m_{\rm{atm}}^{2}$, while simultaneously ensuring dark matter stability. Our model accommodates normal ordering of neutrino masses, with a generalized $\mu$–$\tau$ reflection symmetry shaping the structure of leptonic mixing and a lower limit on the lightest neutrino mass. Moreover, the model provides predictions for the octant of $\theta_{23}$ and a strong correlation between $\Delta m_{\rm sol}^{2}$ and $\Delta m_{\rm atm}^{2}$. This correlation puts a lower bound on the fermionic DM mass.
In contrast, scalar dark matter remains viable over a broad mass spectrum. A notable feature is that the low mass regime ($\sim 15$ GeV onwards) survives owing to the presence of efficient co-annihilation channels, which are typically absent in the Majorana scotogenic scenario. Additionally, the model aligns with current and future limits from lepton flavor violation experiments.

\end{abstract}
\maketitle
%--------------------------------------------------------------------------------------------------------------------------------------------------------------------------------------------------------
%
\section{Introduction}
\label{sec:intro}
The Standard Model (SM) of particle physics has been remarkably successful in explaining the majority of experimental observations at present energy scales. Among the SM particles, neutrinos are the most elusive, as they interact only via the weak interaction. The SM predicts neutrinos to be strictly massless. 
However, the discovery of neutrino oscillations establishes that neutrinos must possess small but nonzero masses \cite{Kamiokande-II:1990wrs,Kamiokande-II:1992hns,Super-Kamiokande:1998kpq,Cleveland:1998nv,SNO:2002tuh}, thereby providing the first clear evidence for physics beyond the Standard Model (BSM).
So far, neutrino experiments have measured two distinct mass-squared differences, $\Delta m_{\rm{sol}}^{2} \approx 10^{-5}~ \text{eV}^2$ and $\Delta m_{\rm{atm}}^{2} \approx 10^{-3} ~\text{eV}^2$, along with one small and two large mixing angles \cite{deSalas:2020pgw,Esteban:2024eli,Capozzi:2025wyn}. The theoretical framework underlying the origin of light neutrino masses and mixing remains one of the persistent challenges in particle physics, commonly referred to as the ``leptonic flavor puzzle''.
Apart from the neutrino sector, another limitation of the SM is the absence of a viable candidate for dark matter (DM). The presence of DM is inferred from a wide range of astrophysical observations and is further supported by cosmological measurements, such as those from the WMAP and Planck missions ~\cite{Zwicky:1933gu,Rubin:1970zza,Rubin:1980zd,Planck:2018vyg}. In view of these observations, DM is generally hypothesized to be a new, electrically neutral, massive particle that interacts only weakly with SM particles. Hence, extensions of the SM that address these issues within a unified framework through new mechanisms offer a promising direction for BSM physics.

In BSM theories, numerous models have been proposed for generating neutrino masses. For these models, Ultraviolet (UV) completions of Weinberg-like operators provide a framework for generating small neutrino masses, either at tree-level or through radiative corrections~\cite{Cai:2017jrq}.
Among the tree-level realizations, the most widely studied are the seesaw frameworks \cite{Schechter:1980gr,Schechter:1981cv,Foot:1988aq,CentellesChulia:2018gwr,Mandal:2023oyh,CentellesChulia:2023osj,Prajapati:2024wuu}.
Beyond tree-level mechanisms, several radiative models \cite{Zee:1980ai,Cheng:1980qt,Babu:2002uu,Krauss:2002px,Ma:2006km}, and their variants have also been widely explored \cite{Guo:2020qin,Dasgupta:2021ggp,Borah:2022enh,Borah:2022phw,ChuliaCentelles:2022ogm,Kumar:2023moh,Kumar:2024zfb,Bharadwaj:2024crt,Borah:2024gql,CentellesChulia:2024iom,Nomura:2024zca,Singh:2025jtn,Kumar:2025aek,Avila:2025qsc,Kumar:2025cte}. Notably, some of these radiative scenarios, such as the scotogenic models, naturally accommodate a stable DM candidate owing to the presence of an imposed $Z_2$ symmetry.
The scotogenic model~\cite{Ma:2006km} has gained significant attention, as it links DM and radiative generation of neutrino masses at the one-loop level. 
Additionally, the scotogenic model can be combined with the seesaw mechanism in a hybrid framework, often referred to as the “scoto-seesaw,” to account for the two distinct neutrino mass scales \cite{Rojas:2018wym}.
Nevertheless, none of these models in their canonical form provides predictions for the neutrino flavor structure.

Among the various proposals to address the leptonic flavor puzzle, one of the most studied frameworks is based on non-Abelian discrete groups, commonly referred to as ``flavor symmetries'' \cite{Babu:1990fr,Frampton:1994rk,Ma:2001dn,Kubo:2003iw,Altarelli:2010gt}. Such symmetries offer a natural explanation of the observed neutrino mixing patterns. Among discrete flavor symmetries, $A_4$ is the most widely explored in the literature~\cite{Ma:2001dn,Babu:2002dz,Chen:2005jm,Altarelli:2005yx}. A notable scotogenic realization of the $A_4$ flavor symmetry was proposed in~\cite{Ma:2008ym}, yielding tri-bimaximal mixing along with the distinctive predictions of the scotogenic model.
In a recent work \cite{Kumar:2024zfb}, the $A_4$ symmetry was employed to the Majorana scotogenic loop to simultaneously explain the observed neutrino mass-squared differences and flavor structure, with its residual $Z_2$ subgroup ensuring DM stability ~\cite{Hirsch:2010ru,Meloni:2010sk,Boucenna:2011tj,Hamada:2014xha,Lamprea:2016egz}. 
In these models, residual $Z_2$ naturally ensures DM stability, unlike in scotogenic models, where the symmetry is imposed by hand.
However, in the typical Majorana scotogenic models, the DM parameter space spans a broad spectrum, extending up to the TeV scale. In contrast, Majorana scotogenic models based on $A_4$ predict a tightly constrained DM mass spectrum~\cite{Boucenna:2011tj,Kumar:2024zfb}. Moreover, in both scenarios,  the low mass DM region is typically excluded by collider and direct detection bounds.

\begin{figure}[h!] 
\centering
       \includegraphics[width=0.6\textwidth]{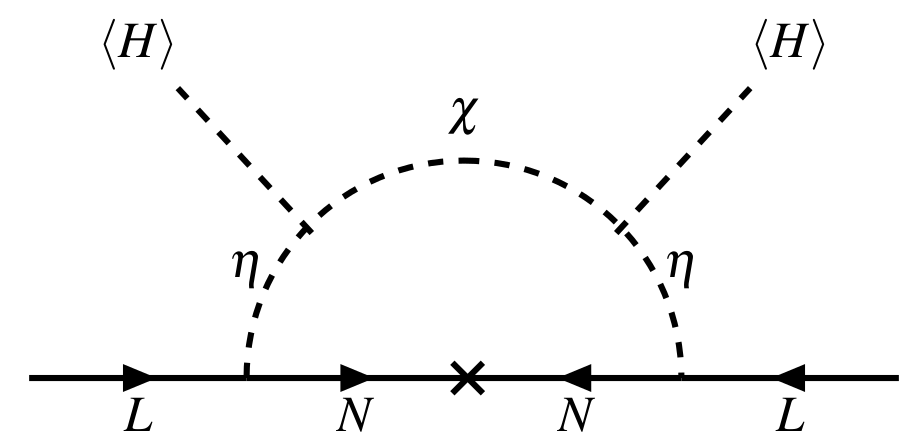}
    \caption{\centering Neutrino mass generation through the scotogenic-like loop.}
    \label{fig:loop1}
\end{figure}

In this work, we propose a variant of the Majorana scotogenic model with $A_4$ symmetry, which accommodates a wide range of DM masses, including the low mass region. Furthermore, this framework simultaneously accounts for neutrino mass generation, the flavor structure, and the two mass-squared differences. We consider a radiative loop involving the SM singlet scalar $\chi$, the scalar $SU(2)_{L}$ doublet $\eta$, and the fermion singlet $N$, as shown in Fig.~\ref{fig:loop1}.
%%%
 The scalars $\eta$ and $\chi$, transforming as $A_4$ triplets, break the $A_4$ symmetry through their vacuum expectation value (VEV). The spontaneous symmetry breaking (SSB) of $A_4$ gives rise to the scoto-seesaw mass mechanism \cite{Rojas:2018wym} and leaves a residual $Z_2$ symmetry that stabilizes the DM candidate ~\cite{Hirsch:2010ru}. 
 The neutrino mass matrix exhibits an interesting flavor structure, leading to the prediction of a generalized $\mu$–$\tau$ reflection symmetry~\cite{Harrison:2002et,Chen:2015siy,Nath:2018hjx,Chakraborty:2018dew,Nath:2018xih,Nath:2018xkz,Nath:2018zoi,Liao:2019qbb,Huang:2020kgt,Yang:2020qsa,Yang:2025yst}. As discussed in Sec.~\ref{sec:numass}, depending on the choice of model parameters, the framework can accommodate either the lower or the higher octant of the atmospheric mixing angle $\theta_{23}$. The model also provides a lower bound on the lightest neutrino mass.
 In addition, the model predicts a robust correlation between $\Delta m^2_{\rm sol}$ and $\Delta m^2_{\rm atm}$. This interplay further places a lower limit on the masses of the BSM fermions, and consequently on the fermionic DM mass.
On the other hand, scalar DM accommodates a broader mass spectrum. This differs from the compressed scalar mass spectrum reported in Ref. \cite{Kumar:2024zfb}, and the difference arises from the presence of the scalar singlet $\chi$ vacuum expectation value (VEV). 
Moreover, scalar $\chi$ introduces additional DM co-annihilation channels. These extra channels make it possible to satisfy the relic density requirement while simultaneously evading collider and direct detection constraints, even in the low mass regime. As a result, viable DM parameter space exists for light and intermediate DM masses, which is typically absent in Majorana scotogenic-like models.
Furthermore, the model successfully satisfies the experimental bounds on lepton flavor violating (LFV) processes, which remain consistent with neutrino oscillation data and DM constraints.
Thus, with a minimal particle content and symmetry, the model simultaneously accounts for neutrino flavor structure and accommodates a wide range of DM masses, notably, the viability of the low mass regime.

The rest of the paper is organized as follows: In Sec.~\ref{sec:model}, we outline the theoretical framework of the model based on $A_4$ symmetry and discuss the Yukawa sector and neutrino mass generation. Sec.~\ref{sec:numass} presents the features of the neutrino mass matrix and its implications for the flavor structure. 
The constraints from LFV processes are discussed in the Sec.~\ref{sec:lfv}.
In Sec.~\ref{sec:dm}, we elaborate on the predictions for the dark sector, with a detailed discussion of the fermionic and scalar DM scenarios. Finally, in Sec.~\ref{sec:conclusion}, we provide concluding remarks.

\section{MODEL FRAMEWORK}
\label{sec:model}
We consider a scotogenic-like loop, where an SM singlet real scalar $\chi$ is inserted between the $SU(2)_L$ doublet scalars $\eta$, as shown in Fig.~\ref{fig:loop1}. The $A_4$ flavor symmetry is employed, which provides a unified framework for neutrino oscillation parameters, DM phenomenology, and testable predictions in LFV processes. In addition to the scalars $\eta$ and $\chi$, a BSM fermion $N$ is introduced, all transforming as $A_4$ triplets. The SM Higgs $H$ transforms as the trivial singlet $(1)$, while the leptons $(L_i, e_{R_i})$ are assigned to the singlet representations $(1,1',1'')$ of $A_4$. The field content and their transformation properties are summarized in Tab.~\ref{tab:fields}.  The transformation properties under residual $Z_2$ symmetry have been listed in the last column. Here, $i = 1, 2, 3$ represents generation indices.
\begin{table}[h]
    \centering
    \begin{tabular}{|c|c||c|c|c|}
        \hline 
     & \  Fields \ & \  $SU(2)_L$ \ & \ $U(1)_Y$ \ & \ $A_4 \to \pmb{Z_2}$ \ \\ \hline \hline
        \ \multirow{3}{*}{\rotatebox{90}{SM}} \ & $L_i$ &  $2$ & $-1$ & $(1,1',1'') \to (\pmb{+,+,+})$  \\
      &  $e_{R_i}$ & $1$ & $-2$  & $(1,1',1'') \to (\pmb{+,+,+})$ \\ 
     &   $H$ & $2$ & $~~1$ & $1 \to \pmb{+}$  \\
        \hline \hline
       \ \multirow{3}{*}{\rotatebox{90}{BSM}} \ &  $\chi$ & $1$ & $~~0$ & $3\to (\pmb{+,-,-})$\\
    &     $\eta$ & $2$ & $~~1$ & $3\to (\pmb{+,-,-})$ \\
     &    $N$ & $1$ & $~~0$ & $3\to (\pmb{+,-,-})$ \\
        \hline
    \end{tabular}
    \caption{Particle content and their transformation properties under different symmetries, including $A_4$ flavor symmetry. Here, $Z_2$ is a residual symmetry arising from $A_4$ symmetry breaking.}
    \label{tab:fields}
\end{table}

The residual $Z_2$ symmetry emerges from the breaking of $A_4$ symmetry through the VEVs of the scalars $\eta$ and $\chi$. 
The specific VEV alignment $\langle \eta \rangle = \tfrac{1}{\sqrt{2}}(v_{\eta},0,0)$ and $\langle \chi \rangle = (v_{\chi},0,0)$ breaks the $A_4$ generator $T$ while leaving $S$ invariant (see App.~\ref{sec:A4}).
Consequently, the $A_{4}$ symmetry is broken down to $Z_{2}$\footnote{Note that Ref. \cite{Ma:2008ym} employs $A_4 \to Z_2$ breaking in a scotogenic framework to obtain tri-bimaximal mixing. However, the DM stability was ensured by the addition of another dark $Z'_2$ symmetry.}
The charge assignments are arranged such that all SM fields, being $A_4$ singlets, remain even under the residual $Z_2$ symmetry. In contrast, the BSM particles, transforming as $A_4$ triplets, can acquire odd $Z_2$ charges, which naturally separates the dark sector from the SM. The even and odd charge assignments under the residual $Z_2$ symmetry are discussed in detail in the App.~\ref{sec:A4}. Under $Z_2$ symmetry, the components of the $A_4$ triplet fields, $\eta_i$, $\chi_i$, and $N_i$ $(i=1,2,3)$, transform as follows:
\begin{equation}
    \begin{aligned}
       & \eta_{1} \rightarrow +\eta_{1},~~\chi_{1} \rightarrow +\chi_{1},~~ N_{1} \rightarrow +N_{1}, \\
       & \eta_{2,3} \rightarrow - \eta_{2,3},~~\chi_{2,3} \rightarrow - \chi_{2,3},~~ N_{2,3} \rightarrow - N_{2,3}\,.
    \end{aligned}
\end{equation}
After symmetry breaking, the particle content can be categorized into two classes based on the residual $Z_{2}$ charge: $Z_{2}$ even and $Z_{2}$ odd particles. The $Z_2$ odd particles belong to the dark sector with the lightest being a potential DM candidate. 
Among the $Z_{2}$ even fields $H$, $\eta_{1}$, and $\chi_{1}$ mix, yielding the Goldstone bosons associated with the $W$ and $Z$ bosons. In addition, we get massive physical particles: one charged scalar $\phi^{\pm}$, one $CP$-odd scalar $A^{0}$, and three $CP$-even scalars $H_i$ ($i=1,2,3$).  
In $Z_2$ odd (dark) sector, $CP$-odd components of $\eta_{k}$ $(k=2,3)$ mix and give rise to two massive physical states, $\eta_{kI}^{0}$. Similarly, the $CP$-even components of $\eta_{k}$ and $\chi_{k}$ mix to yield four massive scalars, denoted by $\zeta_{p}$ $(p=1,2,3,4)$. Also, the mixing between the charged components of $\eta_{k}$ gives two charged scalars $\phi_{k}^{\pm}$.
The scalar sector and their mass spectrum have been discussed in App.~\ref{sec:scalar}.
As mentioned before, the lightest $Z_2$ odd neutral particle among $\eta_{kI}^{0}, \ \zeta_{p}$ and $N_k$ is the stable DM candidate.
For the DM analysis, throughout this paper, we consider either $N_{2}$ as the fermionic DM candidate or $\zeta_{1}$ as the scalar DM candidate.

The breaking of $A_4$ also gives rise to a hybrid scoto-seesaw scenario, where neutrinos acquire their masses from the interplay between the tree-level type-I seesaw and the radiative one-loop scotogenic contribution, as we discuss next.

\subsection{Yukawa Sector and the Generation of Neutrino Mass} \label{sec:yuk}
We now delve into the details of neutrino mass generation. Following the charge assignment given in Tab.~\ref{tab:fields}, the invariant Yukawa Lagrangian that governs the leptonic sector can be expressed as follows:
%%%
\begin{eqnarray}
\label{eq:yuk}
-\mathcal{L}_y &=& y_{e}(\bar{L}_1)_1 H (e_{R_1})_1 +  y_{\mu}(\bar{L}_2)_{1''}H (e_{R_2})_{1'} + y_{\tau}(\bar{L}_3)_{1'}H (e_{R_3})_{1''} + y_1 (\bar{L}_1)_1 \left(\tilde{\eta} N \right)_1 \nonumber \\
&+& y_2 (\bar{L}_2)_{1''} \left(\tilde{\eta} N \right)_{1'} + y_3 (\bar{L}_3)_{1'} \left(\tilde{\eta} N \right)_{1''} + \frac{M}{2} \left(\bar{N}^c N\right)_1 + h.c. \;, 
\end{eqnarray}
where $y_{\alpha}$ ($\alpha=e,\mu,\tau$) and $y_i$ ($i=1,2,3$) are Yukawa couplings that represent the charged lepton and neutrino mass matrix, respectively. $M$ is the Majorana mass of fermion $N$ and $\tilde{\eta}= i \tau_2 \eta^{\ast}$; $\tau_2$ is the second Pauli matrix. The subscript $(\cdots)_{1/1'/1''}$ specifies the $A_4$ representation of the fields, as indicated within the parentheses. The above Lagrangian can further be expressed in terms of the $A_4$ triplet components, using the $A_4$ multiplication rules (see App.~\ref{sec:A4}), as follows:
\begin{align}\label{eq:YukawaLag}
- \mathcal{L}_y & = y_{e}\bar{L}_1 H e_{R_1} +  y_{\mu}\bar{L}_2 H e_{R_2} + y_{\tau}\bar{L}_3 H e_{R_3}  + y_1 \bar{L}_1 \left(\tilde{\eta}_1 N_1 + \tilde{\eta}_2 N_2 +\tilde{\eta}_3 N_3  \right) \nonumber \\ &+ y_2 \bar{L}_2 \left(\tilde{\eta}_1 N_1 + \omega \tilde{\eta}_2 N_2 + \omega^2 \tilde{\eta}_3 N_3  \right) + y_3 \bar{L}_3 \left(\tilde{\eta}_1 N_1 + \omega^2 \tilde{\eta}_2 N_2 + \omega \tilde{\eta}_3 N_3  \right) \nonumber \\ &+ \frac{M}{2}(\bar{N}^c_1 N_1 + \bar{N}^c_2 N_2 + \bar{N}^c_3 N_3) +  h.c. \;.
\end{align}
Here, $\omega$ is the cube root of unity, $\omega^3=1$. 
Once the symmetry is broken, the charged leptons and neutrinos get their masses. The specific $A_4$ charge assignment for the charged leptons ensures that the mass matrix is diagonal, given by:
\begin{align}
    M_{L}= \frac{v_H}{\sqrt{2}} \text{diag} (y_e, y_{\mu}, y_{\tau}) .
\end{align}
Thus, the observed leptonic mixing pattern originates solely from the neutrino sector.

\begin{figure}[!h] 
\centering
  \includegraphics[width=0.85\textwidth]{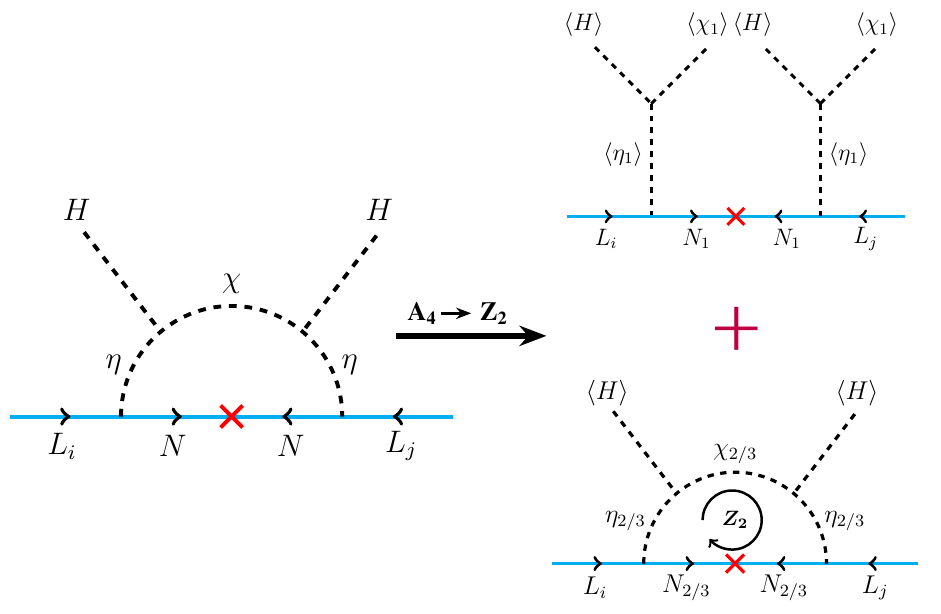}
    \caption{The emergence of the hybrid scoto-seesaw mass mechanism from the breaking $A_4 \to Z_2$ has been shown. The left diagram represents loop before SSB where $A_4$ triplets are running in the loop. After SSB, tree-level seesaw and loop-level scoto diagram have been shown in the right panel. The particles running in the loop (right bottom) are $Z_2$ odd, and the lightest of them serves as a viable DM candidate.}
    \label{fig:A4break}
\end{figure}

After the symmetry breaking, the radiative loop shown in Fig.~\ref{fig:A4break} (left) splits into two parts: tree-level seesaw (top right) and loop-level scotogenic (bottom right), commonly referred to as scoto-seesaw. Thus, neutrinos acquire their masses through this hybrid mass mechanism. 
This arises because, after SSB, the $Z_2$ even scalars $\eta_1$ and $\chi_1$ mix with the Higgs $H$, leading to a type-I like seesaw diagram, with the fermion $N_1$ (being $Z_2$ even) acting as the mediator. The $Z_2$ odd scalars $\eta_{2,3}$ and $\chi_{2,3}$ mix among themselves and run inside a radiative loop together with the $Z_2$ odd fermions $N_{2,3}$. These $Z_2$ odd particles belong to the dark sector, with the lightest among them serving as a viable DM candidate.

Neutrino masses can now be computed from this hybrid mechanism, incorporating contributions at tree-level and loop-level. At the tree-level the mass matrix in the basis of $(\bar{\nu}^c_{L_1},\bar{\nu}^c_{L_2},\bar{\nu}^c_{L_3}, \bar{N}_1^c,\bar{N}_2^c,\bar{N}_3^c)$ and  $(\nu_{L_1},\nu_{L_2},\nu_{L_3},N_1,N_2,N_3)^T$ can be written as
\begin{equation}
\mathcal{M}_{\nu}= \frac{1}{\sqrt{2}}
\begin{pmatrix}
0 & 0 & 0 & y_1 v_{\eta} &0 &0\\
0 & 0 & 0 &y_2 v_{\eta} & 0 &0 \\
0 & 0 & 0 &y_3 v_{\eta}&0 &0\\
y_1 v_{\eta} & y_2 v_{\eta} & y_3 v_{\eta} &\sqrt{2} M &0 &0\\
0 & 0 & 0 &0 &\sqrt{2} M &0\\
0 & 0 & 0 &0 &0 &\sqrt{2} M\\
 \end{pmatrix} \;.
 \label{eq:big-matrix}
\end{equation}
The light neutrino mass matrix in the type-I seesaw limit is given by:
\begin{align} \label{eq:type-i}
    m^{\text{(tree)}}_{\nu} = m_D \mathcal{M}^{-1} m_D^T,
\end{align}
where, matrices $m_D$ and $\mathcal{M}$ are expressed as following:
\begin{align} \label{eq:mat}
    m_D = \frac{1}{\sqrt{2}} \begin{pmatrix}
        y_1 v_{\eta} & 0 & 0\\
         y_2 v_{\eta} & 0 & 0\\
          y_3 v_{\eta} & 0 &0 \\
    \end{pmatrix}, \quad \mathcal{M} =  \begin{pmatrix}
        M & 0 & 0\\
         0 & M & 0\\
          0 & 0 &M \\
          \end{pmatrix}.
\end{align}
Thus, following the Eqs.~\eqref{eq:type-i} and~\eqref{eq:mat}, the light neutrino mass matrix at tree-level is given by\footnote{Due to the $A_4$ symmetry, the masses of BSM fermions ($N_i$) are degenerate and given by $M$. Similar to the type-I seesaw scenario, the mass of $N_1$ receives a small correction and deviates slightly from $M$. At tree-level, $N_2$ and $N_3$ are mass degenerate with mass $M$, but loop effects induce a slight splitting between them.}:
\begin{align} \label{eq:numat}
    m^{\text{(tree)}}_{\nu} = \frac{v^2_{\eta}}{2 M} \begin{pmatrix}
        y_1^2 & y_1 y_2  & y_1 y_3 \\
         y_1 y_2 & y_2^2  & y_2 y_3 \\
          y_1 y_3 & y_2 y_3  & y_3^2 \\
    \end{pmatrix}.
\end{align}
Notably, the rank of matrix $ m^{\text{(tree)}}_{\nu}$ is one, hence only one neutrino acquires mass at tree-level. The other two neutrinos can get their masses through the scoto-loop contribution. 
The light neutrino mass matrix generated at the loop-level can be written as follows (derivation is provided in App.~\ref{sec:loopnumass}):
%%%%    
\begin{align}
m^{(\text{loop})}_{\nu} =
\begin{pmatrix}
y^2_1d'_1 & y_1y_2d'_2 & y_1y_3d'_3  \\
 y_1y_2d'_2 & y^2_2d'_3 & y_2y_3d'_1 \\
y_1y_3d'_3 &  y_2y_3d'_1 & y^2_3d'_2 \\
\end{pmatrix} \;.
\end{align}
%%%%%%%%%%%%%%%%        
% 
\begin{align}\label{eq:dvalues}
\text{where,} \quad \quad d'_1  = c'_2+c'_3, \quad d'_2 =  \omega c'_2+\omega^2c'_3, \quad d'_3 =  \omega^2 c'_2+\omega c'_3.
\end{align}
%%%%%%%%%%%%
Here, $d'_1$ is real, while $\omega$ and $\omega^2$ ensure that $d'_2$ and $d'_3$ are complex conjugates. With real Yukawa couplings, the leptonic CP violation arises exclusively from the complex nature of $d'_2$ and $d'_3$.
By incorporating both the tree-level seesaw and the loop-level scotogenic contributions, the resulting light neutrino mass matrix can be written as
%%%%%%%%%%%%
\begin{equation}\label{eq:NuMassMatrix}
m^{}_{\nu} = m^{(\text{tree})}_{\nu} + m^{(\text{loop})}_{\nu} \equiv
\begin{pmatrix}           
A & C & \tilde{C} \\
C & B & D \\
\tilde{C} & D & \tilde{B} \\                                
 \end{pmatrix} \;.
\end{equation}
%%%%%%%%%%%%%%%  
\begin{align} \label{eq:numatrixpara}
\text{where,} \quad \quad &A =  y^2_1\left(d'_1-\frac{v_{\eta}^2}{2 M}\right),  \quad \quad D =  y_2y_3\left(d'_1-\frac{v_{\eta}^2}{2 M}\right),  \nonumber \\
&B =  y^2_2\left(d'_3-\frac{v_{\eta}^2}{2 M}\right), \quad \quad  \tilde{B} = y^2_3\left(d'_2-\frac{v_{\eta}^2}{2 M}\right), \nonumber \\ 
&C =  y_1y_2\left(d'_2-\frac{v_{\eta}^2}{2 M}\right), \quad  \tilde{C} =  y_1y_3\left(d'_3-\frac{v_{\eta}^2}{2 M}\right).
\end{align}
%%%
Because $d'_1$ is real and the pair $d'_2$, $d'_3$ are complex conjugates, the structure of the mass matrix separates into real and complex components. Specifically, $A$ and $D$ remain real parameters, while $B$, $\tilde{B}$, $C$, and $\tilde{C}$ necessarily take complex values, which in turn provide the source for complex phases in the model. The predictions of this neutrino mass matrix are discussed in the subsequent section. 

\section{Neutrino Sector Predictions} \label{sec:numass}

We now discuss the predictions for the neutrino sector. The model only allows normal ordering (NO) of neutrino masses, while inverted ordering (IO) is disfavored. 
 Our analysis begins with the predictions for the mixing angles and the neutrino mass, highlighting the strong correlations imposed by the model. We then examine how current neutrino oscillation data constrain the allowed parameter space of DM mass, thereby linking neutrino analysis with dark sector phenomenology.

\subsection{Neutrino Flavor Structure and Predictions for the Lightest Neutrino Mass}

 The resulting light neutrino mass matrix in Eq.~\eqref{eq:NuMassMatrix} exhibits the so-called ``generalized $\mu$–$\tau$ reflection symmetry'', a feature that ensures specific predictions for mixing angles and CP phase.
In the particular limit of model parameters, $y_2=y_3$, the elements of $m_{\nu}$ mass matrix reduces as $\tilde{C}= C^{\ast}$ and $\tilde{B}= B^{\ast}$. For this choice $m_{\nu}$ mass matrix becomes $\mu-\tau$ reflection symmetric and predicts $\delta_{\rm{CP}}= \pm \pi/2$ and $\theta_{23} =45^{\circ}$. In this model, the Yukawa couplings $y_2$ and $y_3$ are free parameters and thus the predictions deviate from the exact $\mu$–$\tau$ reflection symmetry. 

\begin{figure}[!h] 
\centering
  \includegraphics[width=0.5\textwidth]{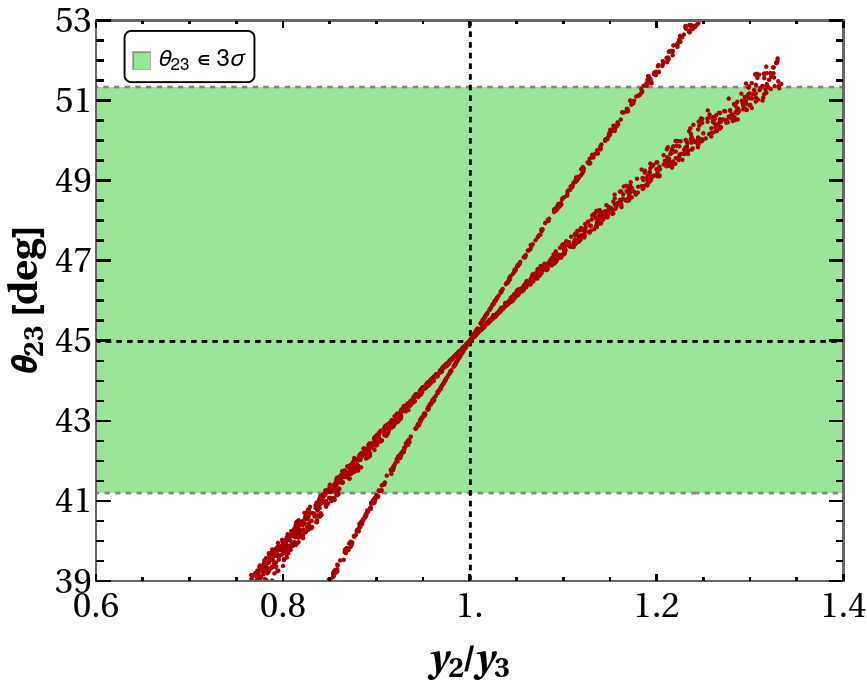}
    \caption{The correlation between atmospheric mixing angle $\theta_{23}$ and ratio of Yukawa couplings $y_2/y_3$.
    }
    \label{fig:th23yuk}
\end{figure}

We begin with the correlations between observables and model parameters. In our analysis, we find that the relative values of Yukawa couplings $y_2, y_3$ determine the octant of the atmospheric mixing angle $\theta_{23}$.
We show this correlation between mixing angle $\theta_{23}$ and the ratio of Yukawa couplings $y_2/y_3$ in Fig.~\ref{fig:th23yuk}. 
The model prediction has been shown in red color points. The green band represents the 3$\sigma$ ranges of the best-fit value of $\theta_{23}$~\cite{deSalas:2020pgw}. The gaps between two bands arise from points excluded by the $3\sigma$ limits on mixing angle $\theta_{13}$, or the two mass-squared differences. From Fig.~\ref{fig:th23yuk}, we see that for $y_2/y_3 = 1$, the mixing angle $\theta_{23}$ takes the value $45^{\circ}$. This has been shown by the intersection of the black dashed line. The condition $y_2 > y_3$ leads to the higher octant, whereas $y_2 < y_3$ favors the lower octant. This relation can be summarized as:
\begin{align}
    &y_2 > y_3 \implies \theta_{23} > 45^{\circ} \quad \quad \quad \text{(higher octant)}, \nonumber \\
    &  y_2 < y_3 \implies \theta_{23} < 45^{\circ} \quad \quad \quad \text{(lower octant)}.
\end{align}

\begin{figure}[!h] 
\centering
  \includegraphics[width=0.45\textwidth]{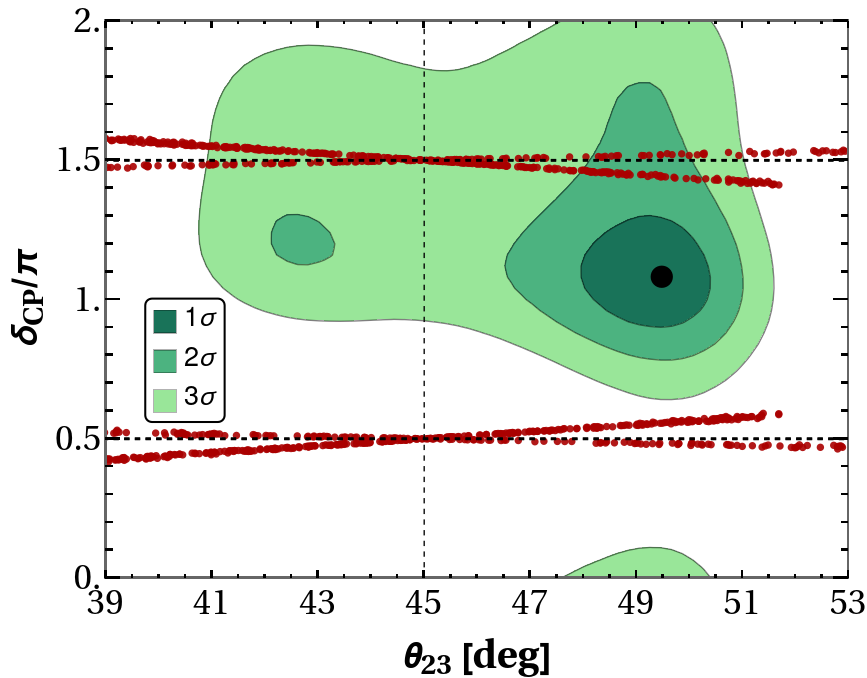}
  \hspace{0.2cm}
  \includegraphics[width=0.45\textwidth]{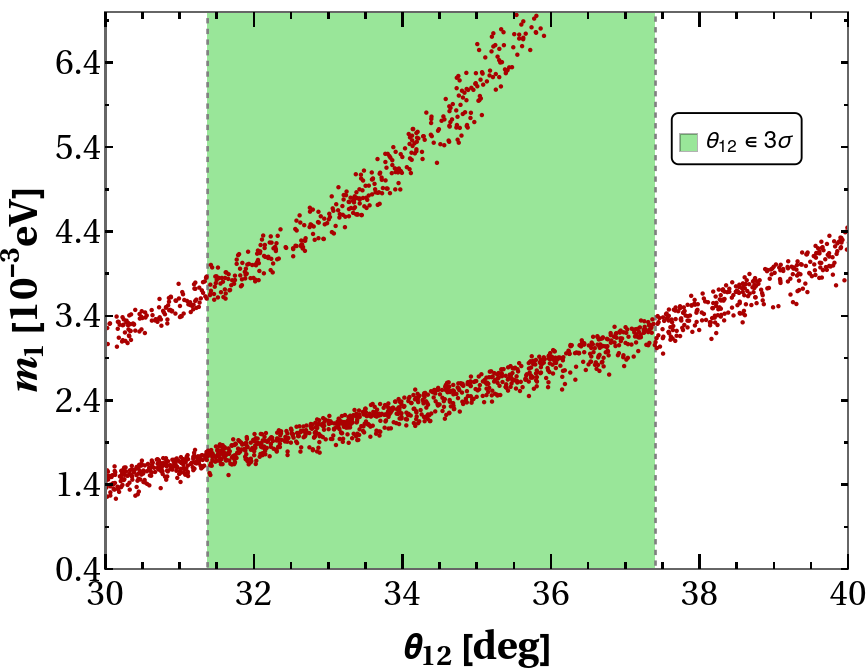}
    \caption{\textbf{Left:} Prediction of generalized $\mu$–$\tau$ reflection symmetry, showing a robust correlation between the CP-violating phase $\delta_{CP}$ and the mixing angle $\theta_{23}$. The intersection of the dashed black lines corresponds to the exact $\mu$–$\tau$ reflection symmetry prediction. \textbf{Right:}  Correlation of the lightest neutrino mass, $m_1 = m_{\rm{lightest}}$ with the mixing angle $\theta_{12}$. The allowed range of $\theta_{12}$ imposes a lower bound on the lightest neutrino mass.}
    \label{fig:delcp}
\end{figure}

We now turn to the model predictions and correlations among the neutrino oscillation observables. These correlations are shown in Fig.~\ref{fig:delcp}. The color scheme and the origin of the gaps between the red points are the same as in Fig.~\ref{fig:th23yuk}, except that the $1\sigma$ and $2\sigma$ contours are also displayed in green shades (left panel). The black dot corresponds to the best-fit value. 
The left panel represents the robust correlation between the CP-violating phase $\delta_{\rm{CP}}$ and $\theta_{23}$. The model predicts that the CP phase is nearly maximally violated ($\delta_{CP} \approx 3 \pi /2$ or $\pi/2$). From this correlation, it follows that only values of $\delta_{CP}$ close to $3 \pi /2$ remain compatible with the latest neutrino data at the $2\sigma$ level. This provides a clear test of the model, since any future measurement of $\delta_{CP}$ substantially deviating from $3 \pi /2$ would rule it out. 
In the right panel of Fig.~\ref{fig:delcp}, we show the strong correlation between the lightest neutrino mass $m_1$ and the mixing angle $\theta_{12}$. We find that imposing the $3\sigma$ range of $\theta_{12}$ puts a lower bound on the lightest neutrino mass. Specifically, the two allowed branches correspond to $m_1 \sim 1.4 \times 10^{-3}$ eV and $m_1 \sim 3.4 \times 10^{-3}$ eV. 
This lower bound on $m_1$ has important implications for neutrinoless double beta decay as well as for cosmology.

Thus, the $A_4$ flavor symmetry not only predicts the octant of $\theta_{23}$ and maximal CP-violating phase but also has implications for the lightest neutrino mass. These model predictions can be probed by experiments such as NOvA, T2K, DUNE~\cite{NOvA:2021nfi,T2K:2023smv,DUNE:2020ypp}, neutrinoless double beta decay experiments~\cite{KamLAND-Zen:2022tow,LEGEND:2021bnm,nEXO:2017nam}, and cosmology~\cite{Planck:2018vyg}.

\subsection{Neutrino Oscillation Constraints on Dark Matter Mass}
\begin{figure}[!h] 
\centering
  \includegraphics[width=0.45\textwidth]{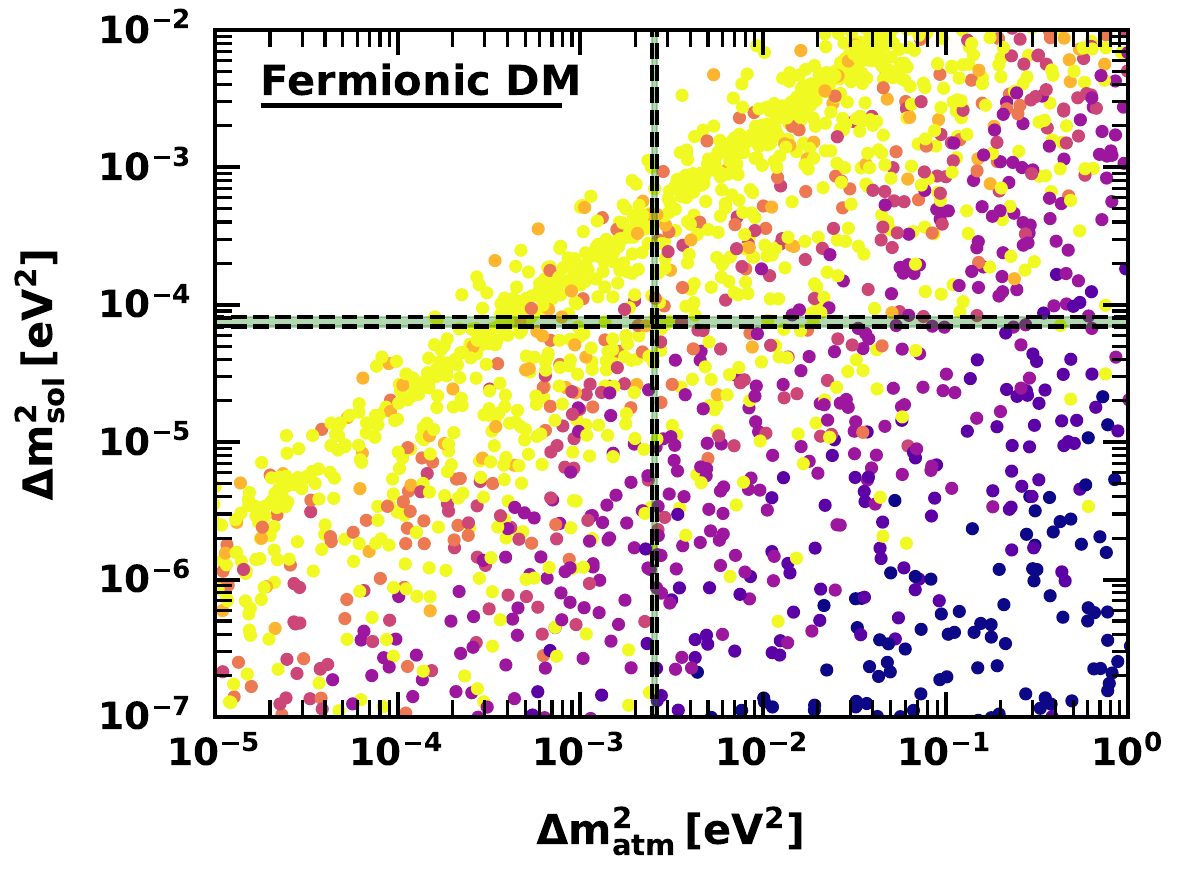}
  \includegraphics[width=0.53\textwidth]{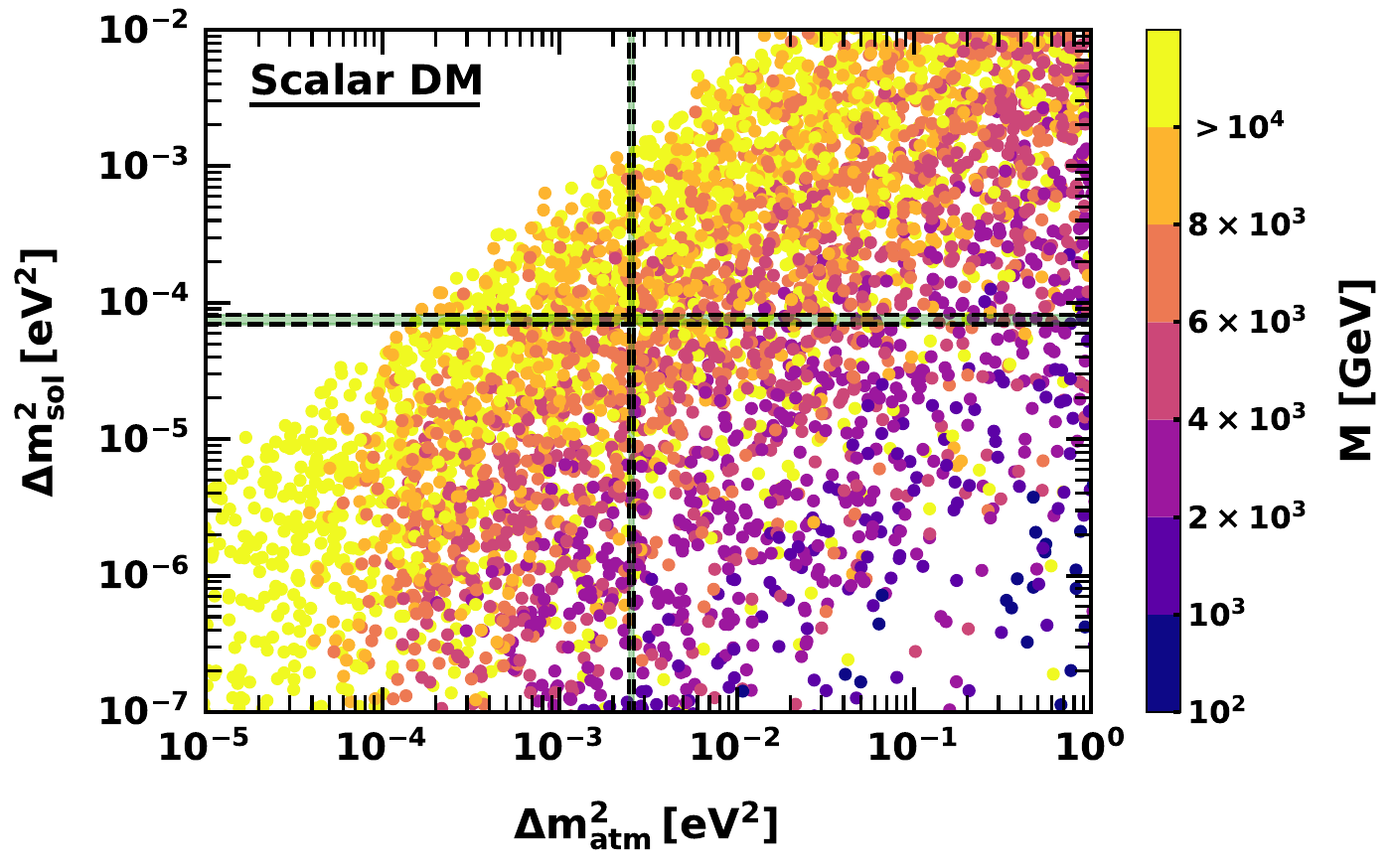}
    \caption{The correlation between two mass-squared differences, $\Delta m^2_{\rm {sol}}$ and $\Delta m^2_{\rm {atm}}$ for the fermionic DM (left) and scalar DM (right) scenarios. The color bar on the right indicates the variation of the fermion mass $M$ for both cases. The yellow region corresponding to $M \gtrsim 10^4$ GeV extends up to the seesaw scale ($10^{12}$ GeV), beyond which the Yukawa couplings reach their perturbativity limit. Only points within the intersection of the vertical and horizontal green bands satisfy both mass-squared difference constraints.}
    \label{fig:ferm_nu}
\end{figure}

In our framework, the $A_4$ symmetry intimately connects the neutrino and dark sectors, thereby shaping the viable parameter space of the model. Neutrino oscillation constraints, in particular, impose a lower bound on the masses of the BSM fermions, which directly determines the scale of the fermionic DM candidate. These constraints not only influence the dark sector phenomenology (see Sec.~\ref{sec:dm}) but also leave imprints on LFV processes (see Sec.~\ref{sec:lfv}). We now present the predictions for the neutrino sector, emphasizing their broader implications for the dark sector.
In Fig.~\ref{fig:ferm_nu}, we show the solar and atmospheric mass-squared differences ($\Delta m^2_{\rm sol}$ and $\Delta m^2_{\rm atm}$) for both the fermionic and scalar DM scenarios. The horizontal and vertical dashed bands represent the 3$\sigma$ ranges of the best-fit values for $\Delta m^2_{\rm sol}$ and $\Delta m^2_{\rm atm}$, respectively~\cite{deSalas:2020pgw}. The color bar on the right indicates the variation of the fermion mass $M$. The model reveals a robust correlation between the solar and atmospheric mass-squared differences.

As a result, the constraint of these two mass-squared differences puts a lower bound on the fermion mass $M$. Only points within the intersection of the vertical and horizontal green bands satisfy both mass-squared difference constraints and are therefore allowed. 
Fig.~\ref{fig:ferm_nu} makes it evident that both $\Delta m^2_{\rm sol}$ and $\Delta m^2_{\rm atm}$ are simultaneously consistent only for sufficiently heavy fermion masses, approximately $ M\gtrsim4$ TeV for both cases. 
This is a reflection of the $A_{4}$ symmetry, which leads to the near-degeneracy of the BSM fermions participating in tree- and loop-level neutrino mass generation. 
Due to this degeneracy, satisfying the two mass-squared differences imposes a lower bound on the fermionic masses.
Thus, the two mass-squared differences observed in neutrino oscillation put a lower bound on the fermion mass $M$ for both fermionic and scalar DM scenarios. This motivates a detailed examination of both the fermionic and scalar DM case, which we discuss in Sec.~\ref{sec:dm}.

 Next, we discuss lepton flavor violation (LFV) and examine how fermion mass constraints affect these processes. We present the LFV analysis first, as it also has implications for the allowed DM parameter space.

\section{Lepton Flavor Violations (LFV)}
\label{sec:lfv}
Many extensions of the SM that generate neutrino masses, particularly low-scale constructions~\cite{Akhmedov:1995ip,Akhmedov:1995vm,Malinsky:2005bi,Mohapatra:1986bd,Gonzalez-Garcia:1988okv}, typically predict sizeable rates for charged lepton flavor violating (cLFV) processes as well. This motivates ongoing and future experimental efforts~\cite{Jodidio:1986mz,MEG:2016leq,COMET:2018auw,Belle:2021ysv,Moritsu:2022lem,Xing:2022rob,MEGII:2023ltw,Perrevoort:2024qtc} dedicated to probing rare flavor violating decays of muons and tau leptons. So far, no definitive evidence for such processes has been reported, and the absence of signals has resulted in stringent upper bounds on the corresponding branching ratios. In the following Tab. \ref{Tab:LFV}, we present the experimental bounds and future sensitivity for different LFV processes.

%%%%%%%%%%%%%%%%%%%%%%%%%%%%%
\begin{table}[h!]
\centering
\begin{tabular}{|c|c|c|}
\hline
\quad LFV decays \quad & Present bound &  \quad Future sensitivity  \quad \\
\hline \hline
\quad BR($\mu \to e \gamma$) \quad & \quad $3.1 \times 10^{-13}$ \cite{MEGII:2023ltw} \quad & $6.0 \times 10^{-14}$ \cite{Baldini:2013ke} \\
\hline
\quad BR($\tau \to e \gamma$) \quad & \quad $3.3 \times 10^{-8}$ \cite{BaBar:2009hkt} \quad & $9.0 \times 10^{-9}$ \cite{Belle-II:2022cgf} \\
\hline
\quad BR($\tau \to \mu \gamma$) \quad & \quad $4.2 \times 10^{-8}$ \cite{Belle:2021ysv} \quad & $6.9 \times 10^{-9}$ \cite{Belle-II:2022cgf} \\
\hline
\end{tabular}
\caption{Present experimental bounds and future sensitivity for different LFV processes.}
\label{Tab:LFV}
\end{table}
%%%%%%%%%%%%%%%%%%%%%%%%%%%%%

\begin{figure}[h!]
\centering
        \includegraphics[height=4.0cm]{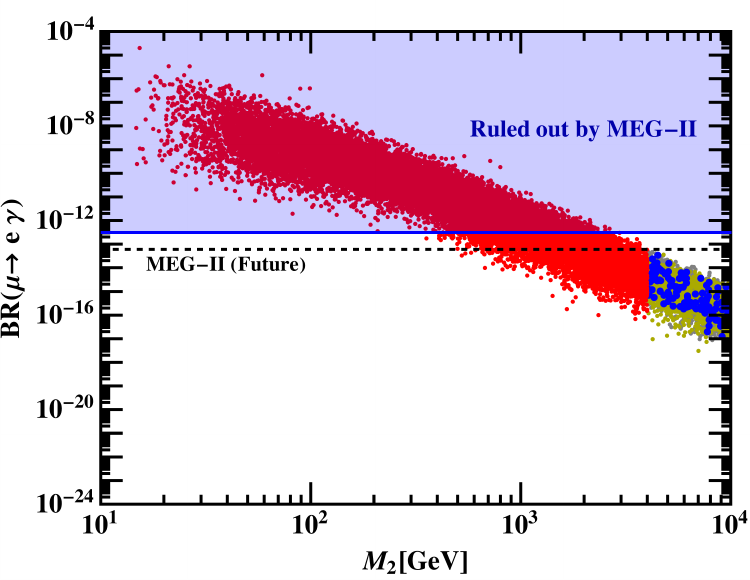}
        \includegraphics[height=4.0cm]{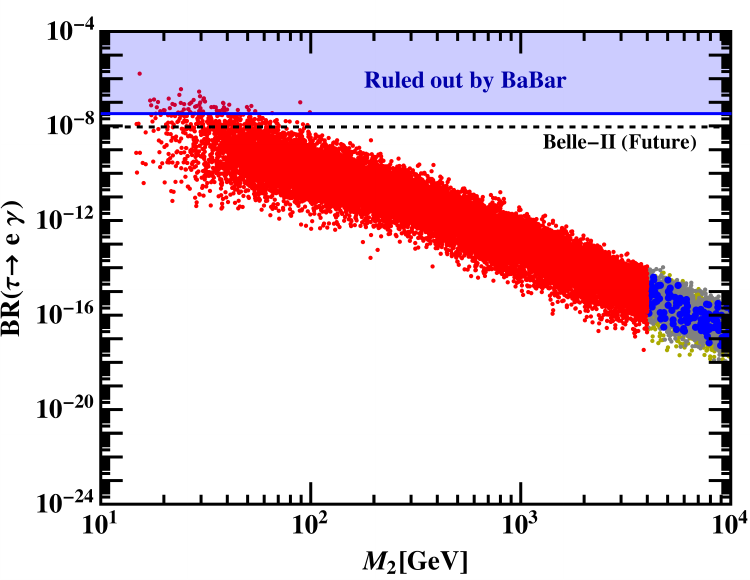}
        \includegraphics[height=4.0cm]{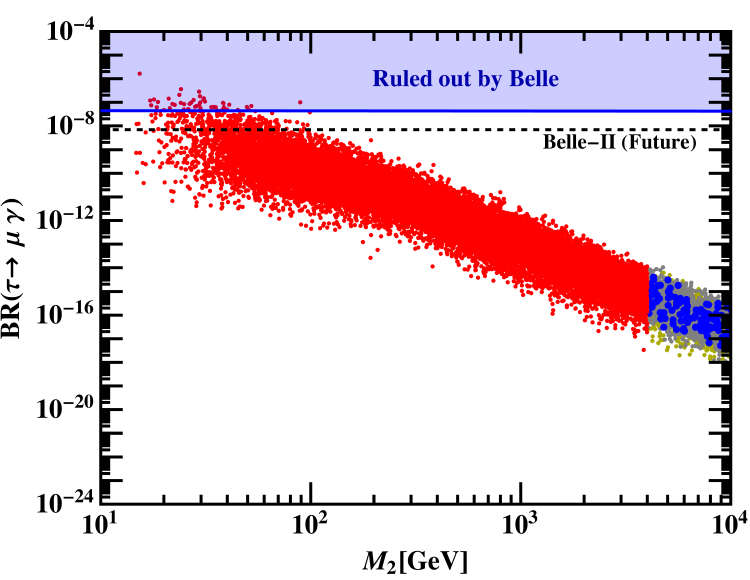}
        \includegraphics[height=4.0cm]{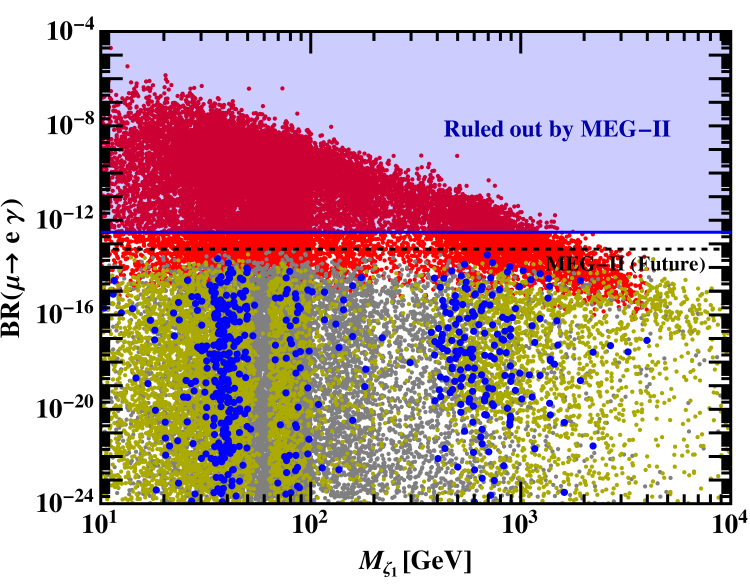}
        \includegraphics[height=4.0cm]{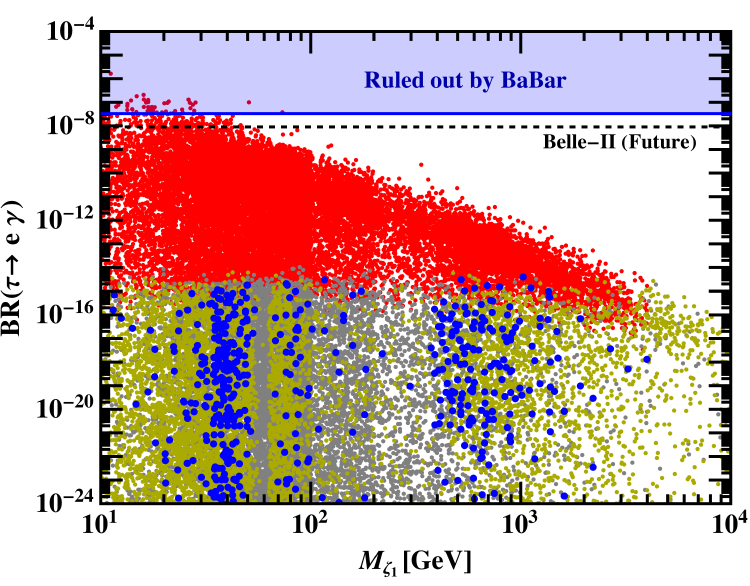}
       \includegraphics[height=4.0cm]{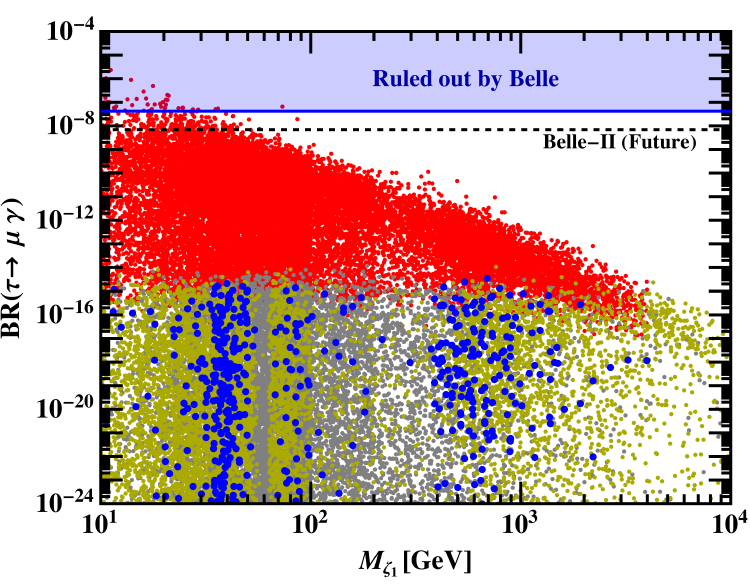}
            \caption{The cLFV rates vs DM mass have been shown. 1st column: BR($\mu \rightarrow e \gamma$), 2nd column:
BR($\tau \rightarrow e \gamma$), 3rd column: BR($\tau \rightarrow \mu \gamma$). 1st row: fermionic DM ($N_2)$ mass, 2nd row: scalar DM ($\zeta_1$) mass. The gray and olive green points represent regions of under abundance and over abundance, respectively, while the blue points correspond to the correct relic density. The red points remain ruled out from neutrino oscillation constraints. 
        }
        \label{fig:LFV}
\end{figure}

We present our model predictions for the various cLFV decay processes shown in Fig.~\ref{fig:LFV}.
The first row of Fig.~\ref{fig:LFV} presents the LFV processes for the fermionic DM case, while the second row corresponds to the scalar DM case. We show the current experimental bounds on cLFV by the purple shaded region, while black dashed lines represent possible future sensitivities. The blue points correspond to the correct relic density, and gray (olive green) represents under (over) abundance of DM. The detailed discussion of fermionic and scalar DM is presented in the next section. The points that fail to satisfy the neutrino oscillation constraints are highlighted in red. 
We also take into account the LHC and LEP constraints, as discussed in the next section. These constraints, however, do not provide any additional constraints on the LFV phenomenology. The LFV is mainly governed by the Yukawa couplings and the masses of the BSM fermions and charged scalars. For this reason, we do not show the invisible constraints explicitly in these LFV plots.
Notably, in both cases, the points that satisfy neutrino oscillation data and the DM relic density lie beyond the current experimental bounds from cLFV searches.
Thus, together, these findings underline robust interplay between the neutrino, DM, and LFV phenomenology, where the viability of neutrino and DM constraints seamlessly translates into compatibility with LFV searches.
In the next section, we present a detailed phenomenological discussion of fermionic and scalar DM scenarios.
%%%%%%%%%%%%%
\section{Dark Sector Phenomenology} \label{sec:dm}

In this section, we analyze the phenomenology of the dark sector within this model. As discussed in Sec.~\ref{sec:model}, once $\eta$ and $\chi$ acquire a VEV, the $A_{4}$ symmetry is spontaneously broken down to a residual $Z_{2}$. All SM fields, together with the BSM fields $\eta_{1}$, $\chi_{1}$, and $N_{1}$, are even under this residual $Z_{2}$ symmetry. In contrast, $\eta_{k}$, $\chi_{k}$, and $N_{k}$ with $k=2,3$ are odd under $Z_{2}$. The $Z_{2}$ even and $Z_{2}$ odd fields mix among themselves, as discussed in App.~\ref{sec:scalar}. After mixing the $Z_2$ odd dark sector consists of four $CP$-even scalars $\zeta_{p}$, two $CP$-odd scalars $\eta_{kI}^{0}$, two charged scalars $\phi_{k}^{\pm}$, and two fermions $N_{k}$ $(k=2,3)$. These fields together define the particle content of the dark sector, whose dynamics are crucial for radiative neutrino mass generation and DM stability. As shown in Fig.~\ref{fig:A4break}, the breaking of $A_{4}$ effectively splits the diagram into two parts: one $Z_{2}$ even and the other $Z_{2}$ odd.
Similar to the scotogenic model~\cite{Ma:2006km}, the $Z_{2}$ odd and electrically neutral fields run inside the loop, with the lightest among them serving as a viable DM candidate. Hence, this model admits both fermionic and scalar DM candidates. 
In our analysis, we consider either the lightest fermion $N_{2}$ or the lightest scalar $\zeta_{1}$ as the DM candidate.
To study the parameter space of these DM candidates, in the following subsection, we discuss the collider constraints arising from invisible decays of the Higgs and $Z$ bosons.

\subsection{Constraints from Higgs and $Z$ Invisible Decays}
In our framework, we have extended scalar sector, making collider constraints particularly important. Interestingly, our analysis shows that the low mass region of the DM parameter space remains viable. In scotogenic models, this regime is typically disfavored because invisible decay searches impose stringent bounds, often excluding the low mass window. Therefore, probing invisible decays plays a crucial role in testing our scenario.
\begin{figure}[!h]
\centering
\includegraphics[height=5.2cm]{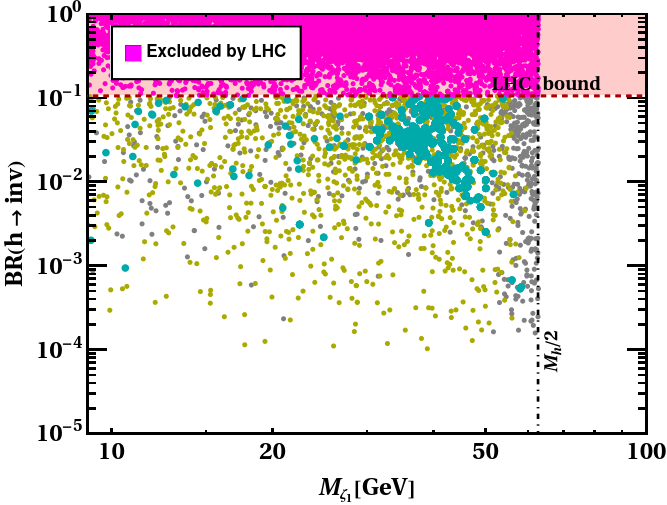}
\includegraphics[height=5.2cm]{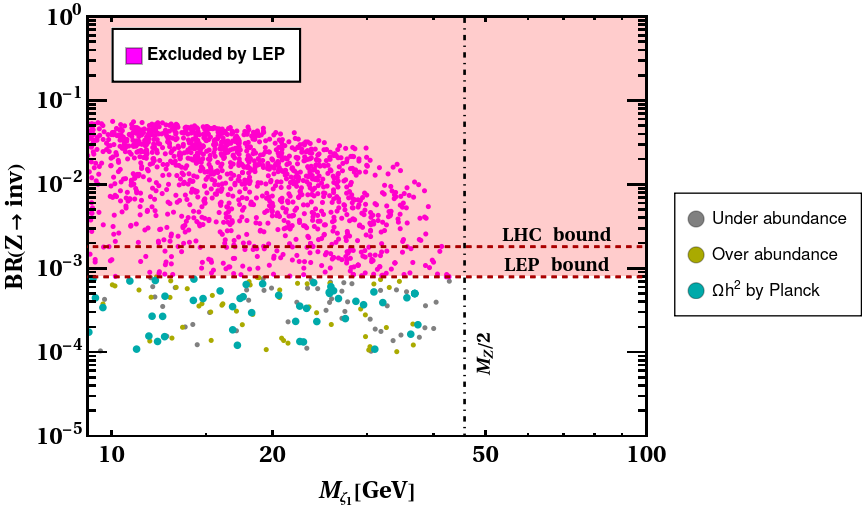}
\caption{Parameter space subject to limits on invisible Higgs decays (left) \cite{ATLAS:2023tkt} and invisible $Z$ decays (right) \cite{LEP:2000pgt,Carena:2003aj,ALEPH:2005ab,ATLAS:2023ynf}. Magenta points are excluded by the corresponding constraint, olive green (gray) points denote over abundant (under abundant) relic density, and cyan points indicate the correct relic density.}
\label{fig:inv}
\end{figure}
Hence, we begin with discussing constraints from invisible decays of Higgs and $Z$ boson derived from LHC ~\cite{ATLAS:2023tkt,ATLAS:2023ynf,ParticleDataGroup:2024cfk} and LEP (LEP-I and LEP-II)  \cite{OPAL:2003wxm,OPAL:2003nhx,LEP:2000pgt,Carena:2003aj,ALEPH:2005ab,Lee:2013fda} data. 
The BSM channel that contributes to invisible Higgs decays includes $h \to A^0 A^0,~h \to \zeta_p \zeta_{p'}$ and $h \to \eta_{kI}^{0} \eta_{k'I}^{0}$. Recent measurement from LHC gives the most stringent constraints on such decays \cite{ATLAS:2023tkt}.
Another constraint comes from $Z$ invisible decays. LEP provides the most precise measurement of the invisible decay width of the $Z$ boson~\cite{OPAL:2003wxm,OPAL:2003nhx,LEP:2000pgt,ALEPH:2005ab,Carena:2003aj}. In the scotogenic type models, the $Z$ boson can decay into pure $SU(2)$ doublet BSM scalars. Such decays are kinematically forbidden to satisfy the LEP bounds, since the decay width is proportional to the $SU(2)$ gauge coupling \cite{Lundstrom:2008ai,Belyaev:2016lok}.
However, in our case, the BSM channel contributing to $Z$ decays includes $Z \to \zeta_{p} \eta_{kI}^{0}$, where $\zeta_{p}$ exhibits a mixed doublet–singlet nature due to the mixing between the doublets $\eta_{2,3}$ and singlets $\chi_{2,3}$. This feature makes the scenario particularly interesting, as the LEP constraints can be somewhat relaxed in this case.
Figure~\ref{fig:inv} shows the constraints on invisible Higgs decays (left panel) and the constraints on invisible $Z$ decays (right panel) from LHC and LEP. The same color scheme is used in both panels: magenta points (and red shaded region) are excluded by the respective constraints, olive green (gray) points correspond to over abundant (under abundant) relic density, and cyan points indicate the correct relic density consistent with the respective constraints.
The vertical dashed lines correspond to half the Higgs mass (left panel) and half the  Z mass (right panel), beyond which the respective decays are kinematically forbidden.

Apart from constraints from invisible Z decays, LEP also imposes bounds on the production of charged scalars~\cite{OPAL:2003wxm,OPAL:2003nhx,Pierce:2007ut} and neutral scalars~\cite{Lundstrom:2008ai}.
In the following subsections, we present a detailed discussion of the parameter space consistent with all the relevant constraints for both fermionic and scalar DM candidates, along with their direct detection prospects.
%%%%%%%%%%%%%
We performed a detailed numerical scan with input parameters given in Tab.~\ref{tab:parameterrange}. 
During the scan, the fermion masses\footnote{As in the usual type-I seesaw scenario, the mass of $N_1$ acquires a small correction and deviates slightly from $M$. The $A_4$ symmetry ensures that $N_2$ and $N_3$ retain their tree-level masses at $M$, although loop effects introduce a slight splitting between them. Taking these small mass differences into account, we denote the fermion masses collectively as $M_i$ for the dark sector analysis.} $M_i$ and Yukawa couplings $y_i$ ($i=1,2,3$) were adjusted to ensure consistency with the neutrino sector analysis. 
\begin{table}[!h]
\begin{center}
\begin{tabular}{| c | c | c | c |}
  \hline 
  Parameter    &   Range   &   Parameter    &   Range  \\
\hline
$\lambda_{H}$     &  	 $[10^{-4},\sqrt{4\pi}]$            &
$\lambda_{\eta_{p}}$   &  $[10^{-8},\sqrt{4\pi}]$            \\
$\lambda_{\chi_{j}}$   &   $[10^{-8},\sqrt{4\pi}]$           &
$|\lambda_{H \eta_{k} }|$   &   $[10^{-8},\sqrt{4\pi}]$   	   \\
$|\lambda_{H \chi_{l}}|$   &   $[10^{-8},\sqrt{4\pi}]$   	   &
$|\lambda_{\eta \chi_{m} }|$   &   $[10^{-8},\sqrt{4\pi}]$   	   \\
$|\kappa_{\chi}|, |\kappa_{\eta\chi}|, |\kappa_{H\eta\chi}|$          &	     $[10^{-8},30]\text{ GeV}$       &
$v_{\chi}$          &	     $[10^{-4},10^{5}]\text{ GeV}$       \\
$M_{{i}}$ & $[10,10^{12}] \text{ GeV}$ 	    &	
$y_{i}$      &  	 $[10^{-7},10^{-1}]$                      \\
    \hline
  \end{tabular}
\end{center}
\caption{The ranges of values used in the numerical scan for the dark sector results correspond to the indices $i,j = 1, 2, 3$,  $k,p = 1, 2, 3, 4, 5$, $l = 1, 2$, and $m = 1, 2, 3, 4$, as defined in the scalar potential given in App.~\ref{sec:scalar}.}
 \label{tab:parameterrange} 
\end{table}

\subsection{Fermionic DM Case}

In this model, the fermion $N_2$ and $N_3$ serve as potential DM candidates, being odd under $Z_2$. Here, we choose $N_2$ as a DM candidate. In our analysis, DM fermion mass $M_{2}$ is taken to be the lightest of all dark sector particles, under the following conditions:
\begin{align}
 M_{2} <  M_{3}, \  M_{\zeta_1}, \ M_{\zeta_2}, \ M_{\zeta_3}, \ M_{\zeta_4}, \ M_{\eta_{2I}^0}, \ M_{\eta_{3I}^0} .
\end{align}
\begin{figure}[!h]
\centering
        \includegraphics[height=7.5cm]{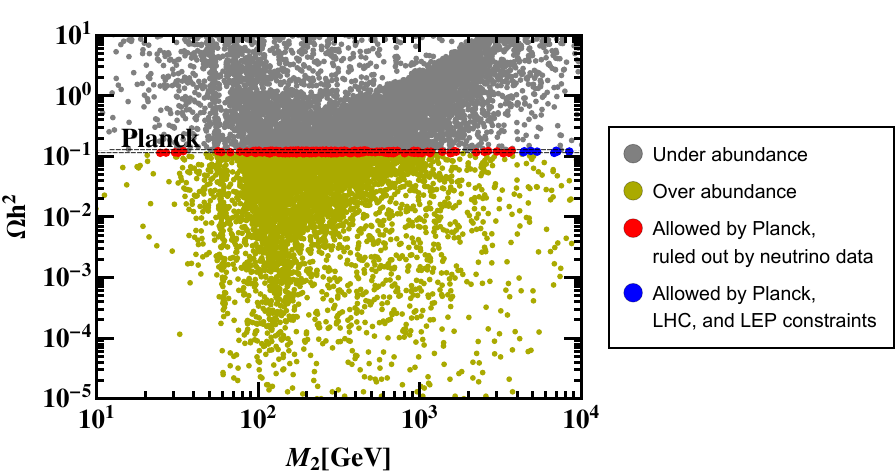}
        \caption{\centering Relic density as a function of fermionic DM mass $M_{N_2}$ has been presented.}
        \label{fig:fermionDM}
\end{figure}
Unlike Ref.~\cite{Kumar:2024zfb}, where the scalar spectrum is confined to masses of around $\sim 600$ GeV, thereby restricting fermionic DM to $\lesssim 600$ GeV, our framework accommodates scalar masses beyond the TeV scale. This, in turn, opens up the possibility of fermionic DM with masses at the TeV scale or higher. In Fig.~\ref{fig:fermionDM}, we present the plot of relic density vs fermionic DM mass. The olive green and gray points correspond to over abundant and under abundant regions, respectively. The blue points lie within the $3\sigma$ range of the relic density derived from Planck satellite data, $0.1126 \leq \Omega  h^2 \leq 0.1246$ \cite{Planck:2018vyg}. The red points are those points which satisfy the relic density but remain ruled out by the neutrino oscillation constraints. 
In addition, we include the constraints from LHC and LEP, shown in Fig.~\ref{fig:inv}, as well as from LFV processes. The LFV bounds are not severe once neutrino oscillation constraints are imposed. In contrast, the invisible channels significantly affect the scalar DM parameter space but have little impact on the fermionic DM case. Therefore, we do not display these constraints explicitly in Fig.~\ref{fig:fermionDM}.
As shown in Fig.~\ref{fig:ferm_nu}, lighter fermion masses fail to satisfy both $\Delta m^2_{\rm sol}$ and $\Delta m^2_{\rm atm}$ simultaneously. 
Therefore, once neutrino oscillation constraints are taken into account, the fermionic DM scenario remains viable only for high mass region, $M_2 \gtrsim 4$ TeV. 

For the fermionic DM scenario, constraints from direct detection experiments such as XENONnT~\cite{XENON:2023cxc,XENON:2025vwd}, LZ~\cite{LZ:2022lsv,LZ:2024zvo}, and PandaX-4T~\cite{PandaX:2024qfu,Zhang:2025ajc} turn out to be rather weak, since the fermionic DM candidate does not couple directly to quarks at tree-level.
However, at the one-loop level, $N_2$ can interact with quarks via the $Z$ boson, photon ($\gamma$), and Higgs boson.
The spin-dependent nucleon cross section is expressed as follows~\cite{Ibarra:2016dlb}:
\begin{align}
\sigma_{SD}\sim 10^{-4} \rm{pb} \left(\frac{\mathit{y_i}}{3}\right)^4 \mathcal{G}_2\left(\frac{\rm{M}^2_2}{m^2_{\eta^{\pm}_{2/3}}}\right)^2,
\end{align}
where, $\mathcal{G}_2(x)$ denotes the loop function, which approaches unity, $\mathcal{G}_2\left(\frac{\rm{M}^2_2}{m^2_{\eta^{\pm}_{2/3}}}\right)^2 \sim 1$, for masses at the TeV scale.
To remain consistent with neutrino oscillation constraints, the Yukawa couplings must be of the order, $y_i\sim 10^{-7} - 10^{-6}$, therefore,
\begin{align} \label{eq:ddcrsec}
\sigma_{SD}\sim 10^{-34} - 10^{-30} \, \rm{pb}.
\end{align} 
Furthermore, the spin-independent cross section per nucleon scales as $\sigma_{SI} \propto y_i^4$, as discussed in~\cite{Ibarra:2016dlb}. Consequently, the spin-independent cross section is likewise negligible.
Since direct detection of fermionic DM arises only at the one-loop level, the resulting spin-dependent (independent) cross sections, $\sigma_{SD}$ ($\sigma_{SI}$), are exceedingly small. As a result, these tiny values easily satisfy the bounds from direct detection experiments~\cite{XENON:2023cxc,XENON:2025vwd,LZ:2022lsv,LZ:2024zvo,PandaX:2024qfu,Zhang:2025ajc}.

\subsection{Scalar DM Case}
The scalar sector, including both the scalar potential and the mass spectrum, is discussed in detail in the App.~\ref{sec:scalar}. The dark sector consist of six scalars $\eta_{kI}^0$ and $\zeta_{p}$ ($k=2,3$ and $p=1,2,3,4$).
The lightest $CP$-odd scalar, $\eta_{kI}^{0}$, is a pure doublet DM candidate, and its phenomenology closely resembles that of conventional doublet DM scenarios.
In the $CP$-even sector, $\zeta_{p}$, exhibits a mixed doublet–singlet nature due to mixing between doublet  $\eta_{2,3}$ and singlet $\chi_{2,3}$ scalars. This feature makes it particularly interesting for further study. Therefore, in this section, we focus on the real scalar DM candidate, namely the lightest among the $\zeta_{p}$.
In our analysis, $\zeta_1$ is taken to be the lightest of all dark sector particles, under the following conditions:
\begin{align}
    M_{\zeta_1} < M_{\zeta_2}, \ M_{\zeta_3}, \ M_{\zeta_4}, \ M_{\eta_{2I}^0}, \ M_{\eta_{3I}^0}, \ M_{2} , \ M_{3} \ .
\end{align}
where, $M_{2}$ and $M_{3}$ are masses of the dark fermion $N_2$, $N_3$. 

\begin{figure}[!h]
\centering
        \includegraphics[height=5.3cm]{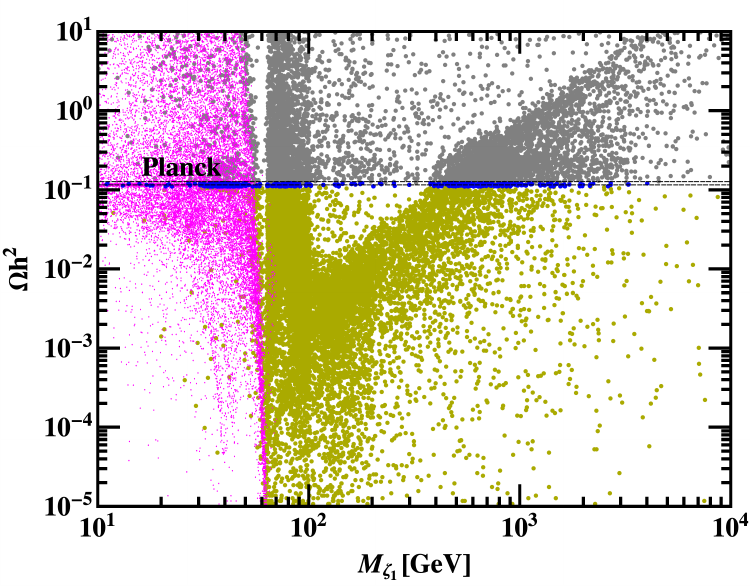}
\includegraphics[height=5.3cm]{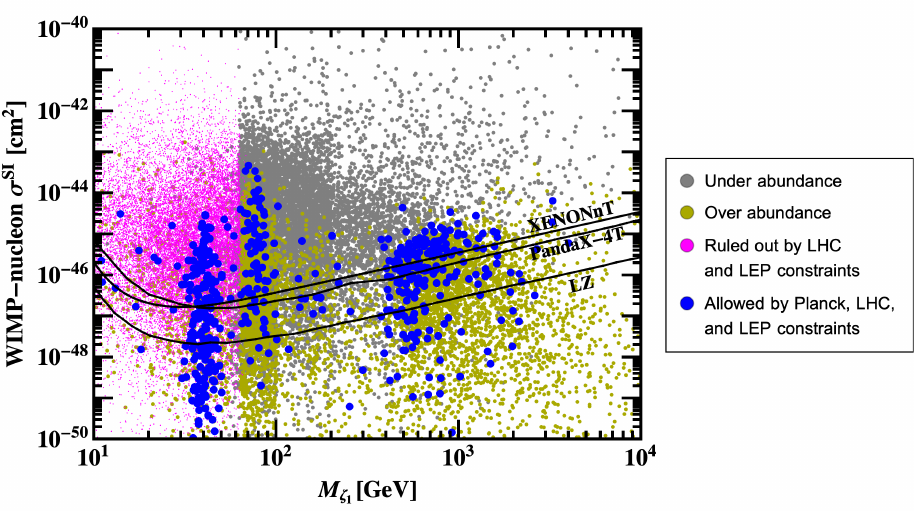}
        \caption{Predictions for the scalar DM candidate. The left panel shows relic density, and the right panel shows WIMP-nucleon cross section vs scalar DM mass.}
        \label{fig:scalarDM}
\end{figure}

The left panel of Fig.~\ref{fig:scalarDM} displays the relic density versus the mass of DM candidate $\zeta_{1}$. The color scheme is the same as in Fig.~\ref{fig:fermionDM}, except for the magenta points, which are ruled out by invisible Higgs decay limits from the LHC~\cite{ATLAS:2023tkt} and LEP constraints~\cite{ATLAS:2023tkt,ParticleDataGroup:2024cfk,OPAL:2003wxm,OPAL:2003nhx,Lundstrom:2008ai,Pierce:2007ut,Carena:2003aj,Lee:2013fda}.
Furthermore, we have taken into account the constraints from neutrino oscillation data and LFV processes, although we do not present them explicitly. The reason is that these constraints do not impose any significant bounds on the scalar DM scenario. Specifically, the constraint from neutrino oscillations primarily sets a lower limit on the BSM fermion masses (see Fig.~\ref{fig:ferm_nu}), which has little impact on the scalar sector due to the presence of sufficient annihilation and co-annihilation channels ensuring the correct relic abundance. Moreover, the LFV bounds derived from neutrino data are already stringent enough (see Sec.~\ref{sec:lfv}), such that all parameter points consistent with neutrino oscillations automatically satisfy LFV constraints.
The right panel of Fig.~\ref{fig:scalarDM} presents the WIMP–nucleon scattering cross section as a function of the DM mass. 
Constraints from recent direct detection experiments such as LZ \cite{LZ:2024zvo}, XENONnT \cite{XENON:2025vwd}, and PandaX-$4T$ \cite{PandaX:2024qfu} are shown by solid black lines, with parameter points lying above these curves excluded by the corresponding experimental limits. Among them, the most stringent bounds are imposed by the LZ result.
From the right panel of Fig.~\ref{fig:scalarDM}, it is evident that the scalar DM exhibits a broad mass spectrum.
Interestingly, the intermediate  $100~\text{GeV} \lesssim M_{\zeta_{1}} \lesssim 300~\text{GeV}$ and the low DM mass range $15~\text{GeV} \lesssim M_{\zeta_{1}} \lesssim 50~\text{GeV}$ remain viable in our framework. These regions are typically excluded in Majorana scotogenic–type models~\cite{Avila:2021mwg,Mandal:2021yph} by collider and direct detection constraints.
This is primarily due to the presence of additional co-annihilation channels in our model. 
Moreover, the higher mass DM region also satisfies both the relic density and direct detection constraints.

%%%%%%%%%%%%%%%%%%%%%%%%
\section{Conclusion}
\label{sec:conclusion}

In this work, we have presented a variant of the Majorana scotogenic model augmented with an $A_4$ flavor symmetry. This framework accommodates a wide range of DM masses, including the low mass region. At the same time, it naturally accounts for neutrino mass generation, the underlying flavor structure, and the two observed mass-squared differences, while ensuring DM stability.
We introduce three BSM fields: an $SU(2)_L$ scalar singlet $\chi$, scalar doublet $\eta$, and a singlet fermion $N$. These BSM fields transform as triplets of $A_4$, while all SM fields remain singlets.
The spontaneous breaking of $A_4$ via the VEVs of the scalar triplets $\eta$ and $\chi$ leads to the emergence of a hybrid scoto-seesaw mechanism and a residual $Z_2$ symmetry, which stabilizes the DM candidate.

Neutrino masses are generated through this hybrid scoto-seesaw mass mechanism that combines the tree-level type-I seesaw with loop-level scotogenic contributions. The neutrino mass matrix exhibits a predictive flavor structure, which naturally realizes a generalized $\mu$–$\tau$ reflection symmetry.  This structure allows the model to accommodate both the lower and higher octants of the atmospheric mixing angle $\theta_{23}$, depending on the choice of Yukawa couplings. These predictions for the flavor structure make the model testable in future neutrino oscillation experiments such as NOvA, T2K, and DUNE. The model predicts the normal ordering of neutrino masses, putting a lower limit on the lightest neutrino mass. Additionally, $\Delta m_{\text{atm}}^2$
and $\Delta m_{\text{sol}} ^2$ have a strong correlation, which also puts a lower limit on the BSM fermion mass. 

Furthermore, the model accommodates both fermionic and scalar DM candidates. To study the DM phenomenology, we have revisited the collider constraints for the LEP and LHC. Unlike the scotogenic models, in the fermionic DM case, the allowed parameter space is pushed into the heavy mass regime $M \gtrsim 4~\text{TeV}$, due to neutrino oscillation constraints. In contrast, scalar DM remains viable over a broad mass regime. Additionally, we find that the presence of the singlet scalar $\chi$ introduces additional DM co-annihilation channels. 
These additional channels open the viable parameter space even in the intermediate, $100~\text{GeV} \lesssim M_{\zeta_{1}} \lesssim 300~\text{GeV}$ and the low mass region,
\(15~\text{GeV} \lesssim M_{\zeta_{1}} \lesssim 50~\text{GeV}\), 
which is typically excluded in Majorana scotogenic-like models.
In addition, the model remains consistent with the current bounds from LFV searches and is compatible with the DM and neutrino oscillation constraints.

In conclusion, this framework offers testable predictions in both the neutrino and dark sectors, which can be probed in forthcoming experiments.

\acknowledgments
\noindent
 SY would like to acknowledge the funding support by the CSIR SRF-NET fellowship. The work of HP is supported by the Prime Minister Research Fellowship (ID: 0401969).

%--------------------------------------------------------------------------------------------------------------------------------------------------------------------------------------------------------
%
\appendix

\section{$A_4$ Symmetry and the Emergence of Residual $Z_2$ Symmetry} \label{sec:A4}
The $A_4$ symmetry is a discrete, non-Abelian flavor group. It is mathematically equivalent to the group of even permutations of four objects, containing 12 elements in total. This group is isomorphic to the symmetry group of a regular tetrahedron. It can be generated by two elements, $S$ and $T$, which obey the following defining relations
\begin{equation} \label{eq:a4genrel}
S^2=T^3=(ST)^3=\mathcal{I}.
\end{equation}
The $A_4$ group consists of four irreducible representations: a triplet (3) and three singlets ($1$, $1'$, and $1''$). The corresponding multiplication rules are given below
\begin{eqnarray} \label{eq:a4mrule}
&1\times1&=1=1' \times 1'', \quad 1'\times 1'=1'', \quad 1''\times 1''=1',  \nonumber \\
&1 \times 3&=3, \quad 3\times 3= 1+ 1' + 1'' + 3_1 + 3_2 \ .
\end{eqnarray}
In a basis where $S$ and $T$ take real matrix representations, they are expressed as
\begin{equation} \label{eq:a4genmat}
S=\left(
\begin{array}{ccc}
1&0&0\\
0&-1&0\\
0&0&-1\\
\end{array}
\right)\,, \quad
T=\left(
\begin{array}{ccc}
0&1&0\\
0&0&1\\
1&0&0\\
\end{array}
\right)\;.
\end{equation}
For two $A_4$ triplets, $x=(x_1,x_2,x_3)$ and $y=(y_1,y_2,y_3)$, the product decomposes according to the following rules~\cite{Ishimori:2010au}
\begin{equation}\label{eq:pr}
\begin{array}{lll}
\left(xy\right)_1&=&x_1y_1+x_2y_2+x_3y_3\, ,\\
\left(xy\right)_{1'}&=&x_1y_1+\omega x_2y_2+\omega^2x_3y_3\, ,\\
\left(xy\right)_{1''}&=&x_1y_1+\omega^2 x_2y_2+\omega x_3y_3\, ,\\
\left(xy\right)_{3_1}&=&\left(x_2y_3,x_3y_1,x_1y_2\right) ,\\
\left(xy\right)_{3_2}&=&\left(x_3y_2,x_1y_3,x_2y_1\right) .
\end{array}
\end{equation}
Here, $\omega^3=1$, denotes a cube root of unity.  
The VEV alignment $\langle \varphi \rangle \sim (v,0,0)$ breaks the generator $T$ and leaves the $S$ generator invariant, thereby breaking the $A_4$ symmetry to its residual $Z_2$ subgroup.
\begin{equation} \label{eq:a4genVEV}
S \langle \varphi \rangle=\langle \varphi \rangle .
\end{equation}
For a $A_4$ triplet field $\psi \equiv (x_1,x_2,x_3)^T$, the action of the generator $S$ is given by the following transformation
\begin{align} \label{eq:a4z2con}
S \psi =\left(
\begin{array}{ccc}
1&0&0\\
0&-1&0\\
0&0&-1\\
\end{array}
\right) \left(
\begin{array}{ccc}
x_1\\
x_2\\
x_3 \\
\end{array}
\right)= \left(
\begin{array}{ccc}
\ \ x_1\\
-x_2\\
-x_3 \\
\end{array}
\right)
\end{align}
 Under $Z_2$ symmetry, the components of the $A_4$ triplet fields, $\eta_i$, $\chi_i$, and $N_i$ $(i=1,2,3)$, transform as follows
\begin{equation}
    \begin{aligned}
       & \eta_{1} \rightarrow +\eta_{1},~~\chi_{1} \rightarrow +\chi_{1},~~ N_{1} \rightarrow +N_{1}, \\
       & \eta_{2,3} \rightarrow - \eta_{2,3},~~\chi_{2,3} \rightarrow - \chi_{2,3},~~ N_{2,3} \rightarrow - N_{2,3}\,.
    \end{aligned}
\end{equation}
All remaining fields in the model remain $Z_2$ even, as they are assigned to the singlet representations of $A_4$. Being singlets, they transform trivially under the action of $S$, ensuring that they remain even under $Z_2$.
\section{The Scalar Sector and Mass Spectrum} \label{sec:scalar}
The scalar sector of the model contains three fields: the SM Higgs doublet $H$, a BSM doublet $\eta$, and a gauge singlet $\chi$. While the SM Higgs is an $A{4}$ singlet, the two BSM scalars $\eta$ and $\chi$ transform as triplets under $A_4$. With this field content, the most general scalar potential invariant under the SM gauge group and $A_4$ can be written as follows:
\begin{align}
    V= V_H + V_{\eta} + V_{\chi} + V_{H \eta} + V_{H \chi} + V_{\eta \chi} + V_{H \eta \chi} + \text{h.c.} \ .
\end{align}
Where,
\begin{equation}
\begin{aligned} \label{eq:pot1}
&V_H=\mu_H^2 H^\dagger H +\lambda_H (H^\dagger H)^2 , \\
&V_{\eta}= \mu_{\eta}^2 (\eta^\dagger\eta)_\mathbf{1} + \lambda_{\eta_{1}}(\eta^\dagger\eta)_{\mathbf{1}}(\eta^\dagger\eta)_{\mathbf{1}}+\lambda_{\eta_{2}}(\eta^\dagger\eta)_{\mathbf{1'}}(\eta^\dagger\eta)_{\mathbf{1''}} \\ &\hspace{1cm}+ \lambda_{\eta_{3}}(\eta^\dagger\eta)_{\mathbf{3_1}}(\eta^\dagger\eta)_{\mathbf{3_1}} + \lambda_{\eta_{4}}(\eta^\dagger\eta)_{\mathbf{3_1}}(\eta^\dagger\eta)_{\mathbf{3_2}} + \lambda_{\eta_{5}}(\eta^\dagger\eta)_{\mathbf{3_2}}(\eta^\dagger\eta)_{\mathbf{3_2}}, \\ 
&V_{\chi}= \mu_{\chi}^2 (\chi \chi)_\mathbf{1} + \lambda_{\chi_{1}}(\chi \chi)_{\mathbf{1}}(\chi \chi)_{\mathbf{1}} + \lambda_{\chi_{2}}(\chi \chi)_{\mathbf{1'}}(\chi \chi)_{\mathbf{1''}} +{\lambda_{\chi_{3}}(\chi \chi)_{\mathbf{3_1}}(\chi \chi)_{\mathbf{3_1}}} + \kappa_{\chi}(\chi)_\mathbf{3} (\chi \chi)_\mathbf{3_1}, \\
&V_{H\eta}= \lambda_{H\eta_{1}} (H^\dagger H)(\eta^\dagger\eta)_\mathbf{1} 
+\lambda_{H\eta_{2}}(H^\dagger\eta)_\mathbf{3}(\eta^\dagger H)_\mathbf{3} +\lambda_{H\eta_{3}}(H^\dagger \eta)_\mathbf{3}(H^\dagger \eta)_\mathbf{3} \\
& \hspace{1cm}+\lambda_{H\eta_{4}}(H^\dagger \eta)_\mathbf{3}(\eta^\dagger \eta)_\mathbf{3_1} + \lambda_{H\eta_{5}}(H^\dagger \eta)_\mathbf{3}(\eta^\dagger \eta)_\mathbf{3_2}, \\
&V_{H\chi}=\lambda_{H\chi_{1}} (H^\dagger H)(\chi \chi)_\mathbf{1} + \lambda_{H\chi_{2}}(H^\dagger\chi)_\mathbf{3}(H \chi)_\mathbf{3}, \\
&V_{\eta \chi}={\lambda_{\eta \chi_{1}}(\eta^\dagger\eta)_{\mathbf{1}}(\chi \chi)_{\mathbf{1}}} + {\lambda_{\eta \chi_{2}}(\eta^\dagger\eta)_{\mathbf{1'}}(\chi \chi)_{\mathbf{1''}}}+ {\lambda_{\eta \chi_{3}}(\eta^\dagger\eta)_{\mathbf{1''}}(\chi \chi)_{\mathbf{1'}}} \\ 
& \hspace{1cm} +\lambda_{\eta \chi_{4}}(\eta^\dagger\eta)_{\mathbf{3_1}}(\chi \chi)_{\mathbf{3_1}} + \kappa_{\eta \chi} (\eta^\dagger \eta)_{\mathbf{3_1}} (\chi )_{\mathbf{3}}, \\
&V_{H\eta \chi}= \lambda_{H \eta \chi } ( H^\dagger \eta)_{\mathbf{3}}(\chi \chi)_{\mathbf{3_1}}  + \kappa_{H \eta \chi} H^\dagger (\eta \chi)_{\mathbf{1}} \, .
\end{aligned}
\end{equation}
The potential is CP conserving if all couplings are taken to be real and  $\lambda_{\eta \chi 2}=\lambda_{\eta \chi 3}$. Now, following the  $A_4$ multiplication rule as given in Eqs.~\eqref{eq:a4mrule} and \eqref{eq:pr}, the expressions given in Eq.~\eqref{eq:pot1} can be simplified in the component form as
%\newpage
%
\begin{align} \label{eq:pot2}
&V_H=\mu_H^2 H^\dagger H +\lambda_H \left(H^\dagger H \right)^2, \nonumber \\
&V_{\eta}= \mu_{\eta}^2 \left(\eta^\dagger_{1}\eta_{1} + \eta^\dagger_{2}\eta_{2} + \eta^\dagger_{3}\eta_{3} \right) + (\lambda_{\eta_{1}} + \lambda_{\eta_{2}})\left((\eta_1^\dagger \eta_1)^2 + (\eta_2^\dagger \eta_2)^2 + (\eta_3^\dagger \eta_3)^2 \right) \nonumber \\
&+ (2 \lambda_{\eta_{1}} - \lambda_{\eta_{2}})\left((\eta^\dagger_{1}\eta_{1})(\eta^\dagger_{2}\eta_{2})  + (\eta^\dagger_{2}\eta_{2})( \eta^\dagger_{3}\eta_{3}) + (\eta^\dagger_{3}\eta_{3} )(\eta^\dagger_{1}\eta_{1}) \right) \nonumber \\
&+ (\lambda_{\eta_{3}} + \lambda_{\eta_{5}}) \left((\eta_1^\dagger \eta_2 )^2 + (\eta_2^\dagger \eta_3 )^2 + (\eta_3^\dagger \eta_1)^2 \right) + \lambda_{\eta_{4}} \left(|\eta_1^\dagger \eta_2|^2 + |\eta_2^\dagger \eta_3|^2 + |\eta_3^\dagger \eta_1|^2 \right) , \nonumber \\
&V_{\chi}= \mu_{\chi}^2 \left(\chi_{1}^2 + \chi_{2}^2 + \chi_{3}^2 \right) + (\lambda_{\chi_{1}} + \lambda_{\chi_{2}})\left(\chi_{1}^4  + \chi_{2}^4 + \chi_{3}^4 \right) \nonumber \\
&+ (2 \lambda_{\chi_{1}}-\lambda_{\chi_{2}} + \lambda_{\chi_{3}})\left( \chi_{1}^2 \chi_{2}^2 + \chi_{2}^2 \chi_{3}^2 + \chi_{3}^2 \chi_{1}^2 \right), \nonumber \\
%%%
&V_{H\eta}= \lambda_{H \eta 1} \left(H^\dagger H \right)\left(\eta_1^\dagger \eta_1 + \eta_2^\dagger \eta_2 + \eta_3^\dagger \eta_3 \right)  + \lambda_{H \eta 2} \left(| H^\dagger \eta_1|^2 + | H^\dagger \eta_2|^2+ |H^\dagger \eta_3|^2  \right) \nonumber \\ 
&+ \lambda_{H \eta 3} \left( H^\dagger \eta_1)^2 + ( H^\dagger \eta_2)^2 + ( H^\dagger \eta_3)^2 \right) + \lambda_{H \eta 4} \left(( H^\dagger \eta_1) (\eta_2^\dagger \eta_3) + ( H^\dagger \eta_2) (\eta_3^\dagger \eta_1) ( H^\dagger \eta_3) (\eta_1^\dagger \eta_2) \right)\nonumber \\ 
&+ \lambda_{H \eta 5} \left(( H^\dagger \eta_1) (\eta_3^\dagger \eta_2) + ( H^\dagger \eta_2) (\eta_1^\dagger \eta_3) ( H^\dagger \eta_3) (\eta_2^\dagger \eta_1) \right), \nonumber \\
&V_{H\chi}= (\lambda_{H\chi_{1}} + \lambda_{H\chi_{2}}) \left(H^\dagger H \right)\left(\chi_1^2 + \chi_2^2 +\chi_3^2  \right), \nonumber \\
&V_{\eta\chi}= (\lambda_{\eta \chi_{1}} + \lambda_{\eta \chi_{2}} + \lambda_{\eta \chi_{3}}) (\eta^\dagger_1 \eta_1 \chi_1^2 + \eta^\dagger_2 \eta_2 \chi_2^2 + \eta^\dagger_3 \eta_3 \chi_3^2) \nonumber \\
&+ \left(\lambda_{\eta \chi_{1}} - \frac{1}{2}\lambda_{\eta \chi_{2}} -\frac{1}{2} \lambda_{\eta \chi_{3}} \right) \left(\eta^\dagger_1 \eta_1 \chi_2^2 + \eta^\dagger_1 \eta_1 \chi_3^2 + \eta^\dagger_2 \eta_2 \chi_3^2 + \eta^\dagger_2 \eta_2 \chi_1^2 + \eta^\dagger_3 \eta_3 \chi_1^2 + \eta^\dagger_3 \eta_3 \chi_2^2 \right), \nonumber \\
&+ \lambda_{\eta \chi_{4}} \left((\eta^\dagger_2 \eta_3)( \chi_2 \chi_3) + (\eta^\dagger_3 \eta_1) (\chi_3 \chi_1) + (\eta^\dagger_1 \eta_2) (\chi_1 \chi_2) \right) + \kappa_{\eta \chi }\left(\eta_1^\dagger \eta_2 \chi_3  + \eta_2^\dagger \eta_3 \chi_1  +\eta_3^\dagger \eta_1 \chi_2  \right), 
\nonumber  \\ 
V_{H \eta \chi}&=  \lambda_{H \eta \chi } \left(( H^\dagger \eta_1) (\chi_3 \chi_2) + ( H^\dagger \eta_2) (\chi_1 \chi_3) + ( H^\dagger \eta_3) (\chi_2 \chi_1) \right)  
\nonumber \\ &  +\kappa_{H \eta \chi }\left(H^\dagger \eta_1 \chi_1  + H^\dagger \eta_2 \chi_2  +H^\dagger \eta_3 \chi_3  \right)\,.
\end{align}
%%%%%%%%%%%%%%%%%%%%%%%%%%%%%%%%%%%%%%%%%%

After the spontaneous breaking of electroweak and $A_4$, the scalar fields can be expanded around minima as,
\begin{align}
%%%%%%%%%%%%%%%%%%%%%%%%%%%%%%%
 H  =
\begin{bmatrix}
H^{+}\\
\frac{1}{\sqrt{2}}(v_H+h+iA)
\end{bmatrix},\, \quad 
    \eta_1 = 
\begin{bmatrix}
\eta_1^{ +}\\
\frac{1}{\sqrt{2}}(v_{\eta}+\eta_{1R}+i\eta_{1I})
\end{bmatrix}, \quad  \chi_1 = v_{\chi} + \chi_{1R}.
%%%%%%%%%%%%%%%%%%%%%%%%%%%%%%%
\end{align}
\begin{align}
%%%%%%%%%%%%%%%%%%%%%%%%%%%%%%% 
    \eta_2 = 
\begin{bmatrix}
\eta_2^{ +}\\
\frac{1}{\sqrt{2}}(\eta_{2R}+i\eta_{2I})
\end{bmatrix}, \quad  \eta_3 = 
\begin{bmatrix}
\eta_3^{ +}\\
\frac{1}{\sqrt{2}}(\eta_{3R}+i\eta_{3I})
\end{bmatrix}, \quad  \chi_2 = \chi_{2R}, \quad  \chi_3 = \chi_{3R} .
%%%%%%%%%%%%%%%%%%%%%%%%%%%%%%%
\end{align}
%%%%%%%%%%%%%%%%%%%%%%%%%%%%%%%%%%%%%%%%%%%%%%%%
The minimization equations for the mass parameters $\mu_{H}, \mu_{\eta}$, and $\mu_{\chi}$ are given by
%%%%%%%%%%%%%%%%%%%%%%%%%%%%%%%%%%%%%%%%%%%%%%%%
\begin{subequations}
\label{tedpole 2}
\begin{align}
&  \mu_{H}^{2} = \lambda_{H} v_H^{2}  + (\lambda_{H \eta 1} + \lambda_{H \eta 2} + 2 \lambda_{H \eta 3}) \frac{v_{\eta}^{2}}{2} + (\lambda_{H \chi 1}+ \lambda_{H \chi 2})v^2_{\chi}+  \kappa_{H \eta \chi} \frac{v_{\chi} v_{\eta}}{v_H} ,\\&
\mu_{\eta}^{2} =  (\lambda_{\eta 1} + \lambda_{\eta 2}  ) v_{\eta}^{2} + (\lambda_{\eta \chi 1} + \lambda_{\eta \chi 2} + \lambda_{\eta \chi 3} ) v_{\chi}^{2} + (\lambda_{H \eta 1} + \lambda_{H \eta 2} + 2 \lambda_{H \eta 3} ) \frac{v_H^{2}}{2}  + \kappa_{H \eta \chi} \frac{v_H v_{\chi}}{v_{\eta}},\\&
\mu_{\chi}^{2} = 2 ( \lambda_{\chi 1} +\lambda_{\chi 2}) v_{\chi}^{2} + (\lambda_{H \chi 1} + \lambda_{H \chi 2} )\frac{v_H^{2}}{2} + (\lambda_{\eta \chi 1} + \lambda_{\eta \chi 12} + \lambda_{\eta \chi 3} ) \frac{v_{\eta}^{2}}{2} +   \kappa_{H \eta \chi} \frac{v_H v_{\eta}}{2 v_{\chi}}  .
\end{align}
\end{subequations}
%%%%

The $Z_2$ even charge scalars $H^{\pm}$ and $\eta_{1}^{\pm}$ will mix and give rise to the Goldstone boson $G^{\pm}$ corresponding to the $W^{\pm}$ boson. One electrically charged field, $\phi^{\pm}$, remains as the physical field. The mass matrix of these charged fields in the basis $(H^{\pm},\, \eta_{1}^{\pm})$ is given by
\begin{equation}
\mathcal{M}_{\pm}^2 =
\begin{pmatrix}
  - (\lambda_{H \eta 2} + 2\lambda_{H \eta 3})\frac{v_{\eta}^2}{2}  - \kappa_{H \eta \chi}\frac{v_{\eta} v_{\chi} }{v_H}  & (\lambda_{H \eta 2} + 2\lambda_{H \eta 3}) \frac{v_H v_{\eta}}{2}  +  \kappa_{H \eta \chi} v_{\chi}  \\
 \ast &  - (\lambda_{H \eta 2} + 2\lambda_{H \eta 3})\frac{v_{H}^2}{2} -\kappa_{H \eta \chi} \frac{v_H v_{\chi}}{v_{\eta}} 
\end{pmatrix}.
\end{equation}
After diagonalizing with a $2\times2$ orthogonal matrix, the physical
mass eigenstate is given as
\begin{equation}
M_{\phi^{\pm}}^{2} = - \left[ ( \lambda_{H \eta 2} +2 \lambda_{H \eta 3})v_H v_{\eta} + \kappa_{H \eta \chi} v_{\chi}  \right] \frac{\left(v_H^{2}+ v_{\eta}^2\right)}{2 v_H v_{\eta}}\,.
\end{equation}
%%%%%%%%%%
The $Z_2$ odd charged scalars $\eta_{2}^{\pm}$ and $\eta_{3}^{\pm}$ mix with each other and after diagonalization with orthogonal matrix give two physical ($ \phi_{2}^{\pm}, \phi_{3}^{\pm}$) fields and resultant mass matrix in the basis $(\eta_{2}^{\pm}, \ \eta_{3}^{\pm})$ is given by
\begin{align}
\mathcal{M}_{\eta_{2/3}^{\pm}}^2=
\begin{pmatrix}
 m_{11}^{\pm}   & m_{12}^{\pm}   \\
  \ast &  m_{22}^{\pm} 
\end{pmatrix}.
\end{align}
The elements of the symmetric mass $\mathcal{M}_{\eta_{2/3}^{\pm}}^2$ can be written as
\begin{align}
& m^{\pm}_{11}= m^{\pm}_{22} =- \frac{3}{2} \lambda_{\eta 2} v^2_{\eta} -\frac{1}{2} (\lambda_{H \eta 2} + 2 \lambda_{H \eta 3} ) v_{H}^{2} -\frac{3}{2}( \lambda_{\eta \chi 2} + \lambda_{\eta \chi 3} ) v^2_{\chi} -\kappa_{H \eta \chi} \frac{v_{H} v_{\chi}}{v_{\eta}}, \\
& m^{\pm}_{12}=\frac{1}{2} (\lambda_{H \eta 4}+\lambda_{H \eta 5} )v_H v_{\eta}+ \kappa_{\eta \chi} v_{\chi}\,.
\end{align}

In the pseudo-scalar sector, $Z_2$ even fields $A$ and $\eta_{1I}$ mix and give Goldstone boson $G^{0}$ corresponding to the neutral gauge bosons $Z$, and one pseudo-scalar field remains as a physical massive field $A^{0}$. The mass matrix in the basis $(A,\eta_{I})$ can be written as
\begin{equation}
\mathcal{M}_{A}^2=
\begin{pmatrix}
- 2 \lambda_{H \eta 3} v_{\eta}^2 - \kappa_{H \eta \chi} \frac{v_{\eta}v_{\chi}}{v_H}   ~&  2  \lambda_{H \eta 3} v_H v_{\eta}  + \kappa_{H \eta \chi} v_{\chi}   \\
\ast ~& - 2 \lambda_{H \eta 3} v_{H}^2 - \kappa_{H \eta \chi} \frac{v_{H}v_{\chi}}{v_{\eta}}
\end{pmatrix}.
\end{equation}
%%%%%%
The mass of the physical eigenstate is given as
\begin{equation}
M_{A^{0}}^{2} = - \left(  2 \lambda_{H \eta 3} v_H v_{\eta}  + \kappa_{H \eta \chi}  v_{\chi}  \right) \frac{\left(v_H^{2}+v_{\eta}^2\right)}{ v_H v_{\eta}}.
\label{eq:mh0}
\end{equation}
The scalars, which are CP-odd as well as $Z_2$ odd, mix with each other and give two physical fields ($\eta_{2 I}^{0},\eta_{3 I}^{0}$). The mass matrix in the basis $(\eta_{2I}, \ \eta_{3I})$ is given by
\begin{align}
\mathcal{M}_{\text{I}}^2=
\begin{pmatrix}
 m_{I11} & m_{I12}   \\
  \ast &  m_{I22} 
\end{pmatrix}.
\end{align}
The elements of the symmetric mass $\mathcal{M}_{\text{I}}^2$ can be written as
\begin{align}
& m_{I11}= m_{I22} =- \frac{1}{2} (3 \lambda_{\eta 2} + 2 \lambda_{\eta 3} -\lambda_{\eta 4} + 2 \lambda_{\eta 5} ) v_{\eta}^{2} -\frac{3}{2}( \lambda_{\eta \chi 2} + \lambda_{\eta \chi 3} ) v^2_{\chi} - \kappa_{H \eta \chi}\frac{v_{H} v_{\chi}}{v_{\eta}}, \\
& m_{I12}=\frac{1}{2} (\lambda_{H \eta 4}+\lambda_{H \eta 5} )v_H v_{\eta}+ \kappa_{\eta \chi} v_{\chi}\,.
\end{align}
The mass eigen states $\eta_{j I}^{0}$ are related to the gauge fields $\eta_{j I}$ through an orthogonal rotation,  
$\eta_{j I}^{0} = \mathcal{O}_{I}[j,k-1]\,\eta_{k I}$,  
where $j,k = 2,3$, and $\mathcal{O}_{I}$ denotes an orthogonal rotation matrix.

Three CP and $Z_2$ even neutral scalars are mixed together. The mass matrix in the basis $(h,\eta_{1R},\chi_{1R})$ can be expressed as
\begin{align}
\mathcal{M}_{H}^2=
\begin{pmatrix}
2 \lambda_{H} v_H^{2}- \kappa_{H \eta \chi} \frac{v_{\eta} v_{\chi}}{v_H}  & m_{H12}  &  2 (\lambda_{H \chi 1} +\lambda_{H \chi 2}) v_H v_{\chi} + \kappa_{H \eta \chi} v_{\eta} \\\\
  \ast & 2( \lambda_{\eta 1} + \lambda_{\eta 2}) v_{\eta}^{2}-\kappa_{H \eta \chi}\frac{v_H v_{\chi}}{v_{\eta}} & 2 (\lambda_{\eta \chi 1} + \lambda_{\eta \chi 2} + \lambda_{\eta \chi 3}) v_{\eta} v_{\chi} + \kappa_{H \eta \chi} v_H \\\\
  \ast & \ast & 8(\lambda_{\chi 1} + \lambda_{\chi 2}) v_{\chi}^{2}- \kappa_{H \eta \chi} \frac{v_H v_{\eta}}{v_{\chi}} 
\end{pmatrix}.
\end{align}
%%%%%%%%%%%%%
where $m_{H12}=\left( \lambda_{H \eta 1}+\lambda_{H \eta 2}+2\lambda_{H \eta 3} \right)v_H v_{\eta} + \kappa_{H \eta \chi} v_{\chi}$.
%%%%%%%%%%%%%
This matrix is diagonalized by an $3\times 3$ orthogonal matrix $\mathcal{O}_{H}$. After diagonalization, the mass eigenstates are $(H_{1},H_{2},H_{3})$.    
%%%
We assume that the mass eigenstates are ordered as $ M_{H_1} < M_{H_2} < M_{H_3}$. The lightest mass eigenstate $H_1 \equiv h$ is identified as the SM Higgs of mass 125 GeV.
%%%%%%%%%%%%%

The four scalar fields that are CP-even but $Z_{2}$ odd mix with each other to yield four physical states. The mass matrix in the basis $(\eta_{2R},\, \eta_{3R},\, \chi_{2R},\, \chi_{3R})$ is given by
\begin{align}
\mathcal{M}_{\text{R}}^2=
\begin{pmatrix}
 m_{R11}   & m_{R12}  & m_{R13} & \lambda_{H \eta \chi} v_{H} v_{\chi} + \kappa_{ \eta \chi} v_{\eta} \\\\
  \ast &  m_{R22}  & \lambda_{H\eta \chi } v_{H} v_{\chi} + \kappa_{\eta \chi} v_{\eta} \ &  \ \lambda_{\eta \chi 4} v_{\eta} v_{\chi} + \kappa_{H \eta \chi} v_{H}\\\\
  \ast & \ast &  m_{R33}& \lambda_{H \eta \chi} v_{H} v_{\eta} + \kappa_{\chi} v_{\chi} \\\\
 \ast & \ast & \ast & m_{R44}
\end{pmatrix}.
\end{align}
Where,
\begin{align}
& m_{R11}= m_{R22} =\frac{1}{2} (-3 \lambda_{\eta 2} + 2 \lambda_{\eta 3} + \lambda_{\eta 4} + 2 \lambda_{\eta 5} ) v_{\eta}^{2}  -\frac{3}{2}( \lambda_{\eta \chi 2} + \lambda_{\eta \chi 3} ) v^2_{\chi} - \kappa_{H \eta \chi} \frac{v_{H} v_{\chi}}{v_{\eta}}, \\
& m_{R12}=\frac{3}{2} (\lambda_{H \eta 4}+\lambda_{H \eta 5} )v_H v_{\eta}+ \kappa_{\eta \chi} v_{\chi},~ m_{R13}=  \lambda_{\eta \chi 4} v_{\eta} v_{\chi} + \kappa_{H \eta \chi} v_{H},\\
& m_{R33}= m_{R44} = 2( \lambda_{\chi 3} - 3 \lambda_{\chi 2} ) v^2_{\chi} -\frac{3}{2}( \lambda_{\eta \chi 2} + \lambda_{\eta \chi 3} ) v^2_{\eta} - \kappa_{H \eta \chi} \frac{v_H v_{\eta}}{v_{\chi}}.
\end{align}
The mass eigenstates $\zeta_{p}$ are obtained from the gauge eigenstates $\eta_{R} = (\eta_{2R},\, \eta_{3R},\, \chi_{2R},\, \chi_{3R})$ through the orthogonal transformation 
$\zeta_{p} = \mathcal{O}_{R}[p,q]\,\eta_{R_{q}}$,  
with $p,q = 1,2,3,4$.
%
%%%%%%%%%%%%%%%%%%%%%%%%%%%%%%%%%%%%%%%%%%
\section{Calculation of Neutrino Mass Generation at Loop-Level} \label{sec:loopnumass}
\begin{figure}[!h] 
\centering
\includegraphics[width=0.49\textwidth]{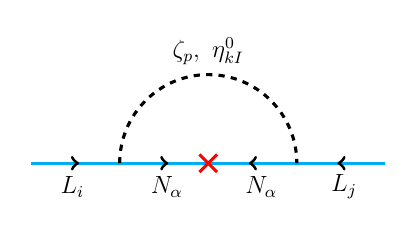}
    \caption{One-loop induced neutrino mass diagram, where $Z_{2}$ odd dark sector scalars and fermions propagate in the loop and generate Majorana masses for the light neutrinos.}
    \label{Fig:Loop_appn}
\end{figure}

Here, we present the calculation of the one-loop induced neutrino mass\footnote{In this analysis, we neglect the one-loop correction to the type-I seesaw, as it is subdominant compared to the tree-level contribution to the neutrino mass.}. The Feynman diagram corresponding to the loop-level scoto contribution is shown in Fig.~\ref{Fig:Loop_appn}. In this case, the $Z_{2}$ odd dark sector fields: the real scalars $\zeta_{p}$, the pseudo-scalars $\eta_{kI}^{0}$, and the fermion $N_{\alpha}$, propagate in the loop.

\noindent
 The resulting one-loop contribution to the neutrino mass from these particles is given by
\begin{align} \label{eq:loopmass}
  m_{\nu}^{\text(loop)}=  \sum^3_{i, j =1} \left(M_{\nu}\right)_{i j}=  \sum^3_{i, j =1} \left(Y_{i \alpha} Y_{j \alpha} c_{\alpha}^{1} + Y_{i \alpha} Y_{j \alpha} c_{\alpha}^{2} \right), 
\end{align}
where $\alpha=2,3$ and $c_\alpha^1$, $c_\alpha^2$ are loop contributions given by:
\begin{align}
   & c_{\alpha}^{1} = \frac{M}{32 \pi^{2}} \left[ (\mathcal{O}_{R}^{T}[\alpha-1, p])^2 \mathcal{F} \left(M_{\zeta_{p}} \right)
   \right],~~p=1,2,3,4~~,\\
      & c_{\alpha}^{2} = -\frac{M}{32 \pi^{2}} \left[ (\mathcal{O}_{I}^{T}[\alpha-1, k-1])^2 \mathcal{F} \left(M_{\eta_{kI}^{0}} \right)
   \right],~~k=2,3~~.
\end{align}
Here, $\mathcal{F} \left(m_i \right)$ denotes the loop function, defined as,
\begin{align}
 \mathcal{F}\left(m_i \right)= \frac{m_i^2}{m_i^2 - M^2} \ln{\left( \frac{m_i^2}{M^2} \right)}.
\end{align}
The factors $Y_{i\alpha}$ and $Y_{j\alpha}$ appearing in Eq.~\eqref{eq:loopmass} represent the effective Yukawa couplings generated at the one-loop level. Their explicit forms are given by
\begin{equation}
\begin{aligned}
    & Y_{11} = y_{1},~Y_{12} = y_{1},~Y_{13}=y_{1},  \\ 
    &Y_{21}=y_{2},~Y_{22}= \omega y_{2},~ Y_{23} = \omega^{2} y_{2},  \\ 
    &Y_{31}= y_{3},~Y_{32}= \omega^{2} y_{3},~Y_{33} = \omega y_{3}\,.
\end{aligned}
\label{Eq:Loop_Yukawa}
\end{equation}
Using Eqs.~\eqref{eq:loopmass} and \eqref{Eq:Loop_Yukawa}, the elements of the neutrino mass matrix arising at the one-loop level can be expressed as,
\begin{equation}
\begin{aligned}
   & (M_{\nu})_{11} = y_{1}^{2} \left[c'_{2} + c'_{3}\right],~ (M_{\nu})_{12} = y_{1} y_{2} \left[c'_{2} \omega + c'_{3} \omega^{2} \right],~(M_{\nu})_{13} = y_{1} y_{3} \left[ c'_{2} \omega^{2} + c'_{3} \omega \right]\,,  \\
   & (M_{\nu})_{21} = y_{1} y_{2} \left[ c'_{2} \omega + c'_{3} \omega^{2} \right],~(M_{\nu})_{22} = y_{2}^{2} \left[ c'_{2} \omega^{2} + c'_{3} \omega \right],~ (M_{\nu})_{23} = y_{2} y_{3} \left[c'_{2} + c'_{3} \right]\,,  \\
   & (M_{\nu})_{31} = y_{1}y_{3} \left[c'_{2} \omega^{2} + c'_{3} \omega \right], ~ (M_{\nu})_{32} = y_{2}y_{3} \left[c'_{2} \omega^{2} + c'_{3} \omega \right],~(M_{\nu})_{33} = y_{3}^{2} \left[c'_{2} \omega + c'_{3} \omega^{2} \right]\,.  
\end{aligned}
\end{equation}
Where $c'_2 \equiv c_2^1 + c_2^2$ and $c'_3 \equiv c_3^1 + c_3^2$.
Hence, the light neutrino mass matrix generated at the one-loop level can be written as follows:
%%%%    
\begin{align}
m^{(\text{loop})}_{\nu} =
\begin{pmatrix}
y^2_1d'_1 & y_1y_2d'_2 & y_1y_3d'_3  \\
 y_1y_2d'_2 & y^2_2d'_3 & y_2y_3d'_1 \\
y_1y_3d'_3 &  y_2y_3d'_1 & y^2_3d'_2 \\
\end{pmatrix} \,,
\end{align}
%%%%%%%%%%%%%%%%        
% 
\begin{align}\label{eq:dvalues1}
\text{where,} \quad \quad d'_1  = c'_2+c'_3, \quad d'_2 =  \omega c'_2+\omega^2c'_3, \quad d'_3 =  \omega^2 c'_2+\omega c'_3 \ .
\end{align}
%%%%%%%%%%%%
Here, $d'_1$ is real, while $\omega$ and $\omega^2$ ensure that $d'_2$ and $d'_3$ are complex conjugates. 
\bibliographystyle{utphys}
\bibliography{references}
\end{document}